\documentclass[11pt,a4paper,twoside]{article}

\usepackage{blindtext} 

\usepackage[utf8]{inputenc}
\usepackage[T1]{fontenc}
\usepackage[english]{babel}

\usepackage[top=2cm, bottom=2cm, outer=2cm, inner=2cm]{geometry}
\usepackage{fancyhdr}	
\usepackage{titlesec}

\usepackage{lettrine} 

\usepackage{enumitem} 
\setlist[itemize]{noitemsep} 

\usepackage{abstract} 

\usepackage{titlesec} 
\titleformat{\section}[block]{\Large\bfseries}{\thesection.}{1em}{} 
\titleformat{\subsection}[block]{\large\bfseries}{\thesubsection.}{1em}{} 

\usepackage{fancyhdr} 
\usepackage{titling} 

\usepackage{hyperref} 

\usepackage{etoolbox}

\usepackage{titletoc}
\usepackage{minitoc}

\usepackage{setspace}   
\usepackage{soul}           
\usepackage[11pt]{moresize}
\usepackage{textcomp}  
\usepackage{lettrine}
\usepackage{lmodern}  
\usepackage{microtype}	
\usepackage{url}
\usepackage{nth}
\usepackage{fancybox}
\usepackage{footmisc}     
\usepackage{epigraph}
\usepackage{multirow}	
\usepackage{tabularx}	
\usepackage[normalem]{ulem} 

\usepackage{color}
\usepackage[usenames, dvipsnames]{xcolor}
\definecolor{bleuuniv}{RGB}{0,54,103}
\definecolor{vertuniv}{RGB}{212,215,0}
\usepackage{graphicx}
\usepackage{caption}
\usepackage{xkeyval}
\usepackage{subfig}
\usepackage{tabularx}
\usepackage{wrapfig}
\usepackage{tikz}
	\usetikzlibrary{patterns,arrows,decorations.pathmorphing}
\usepackage{tikz-3dplot}
\usepackage{float}
\usepackage{tcolorbox}

\usepackage{amsmath}  
\usepackage{amssymb}    
\usepackage{mathrsfs}   
\usepackage{mathtools}	
\usepackage{braket}   
\usepackage[autolanguage]{numprint}   
\usepackage{isotope}   
\usepackage{listings}   
\usepackage{verbatim}  
\usepackage{fancyvrb}   
\usepackage{wasysym}		
\usepackage[load-configurations = abbreviations,alsoload=hep]{siunitx}
\DeclareSIUnit{\sample}{S}
\usepackage{engord}

\pretitle{\begin{center}\huge\bfseries} 
\posttitle{\end{center}} 
\title{Radio detection of cosmic rays in [1.7-3.7]~MHz: the EXTASIS experiment} 
\author{\small D. Charrier$^{1,2}$, R. Dallier$^{1,2}$, Antony Escudie$^1$, D. Garc\'{\i}a-Fern\'{a}ndez$^1$, A. Lecacheux$^3$, L. Martin$^{1,2}$\\ and B. Revenu$^{1,2}$ \\
        \small $^1$ SUBATECH, Institut Mines-Telecom Atlantique -- CNRS/IN2P3 -- Universit\'e de Nantes, Nantes, France \\
        \small $^2$ Unit\'e Scientifique de Nan\c cay, Observatoire de Paris, CNRS, PSL, UO/OSUC, Nan\c cay, France \\
        \small $^3$ CNRS-Observatoire de Paris, Meudon, France}
        
\date{\today} 
\renewcommand{\maketitlehookd}{%
\begin{abstract}
\noindent Since 2003, significant efforts have been devoted to the understanding of the radio emission of extensive air showers above \SI{20}{\mega\hertz}. Despite some studies led until the early nineties, the band available below \SI{20}{\mega\hertz} has remained unused for 20 years. However, it has been claimed by some pioneering experiments that extensive air showers emit a strong electric field in this band and that there is evidence of a large increase of the radio pulse amplitude with decreasing frequencies. The EXTASIS experiment, located within the Nan\c{c}ay Radioastronomy Observatory and supported by the scintillator array and the autonomous radio stations of the CODALEMA experiment, aims to re-investigate the low-frequency band, and especially to study the so-called "sudden death" contribution, the expected electric field radiated by the shower front when hitting ground level. In this work, we present the instrumental setup of the EXTASIS experiment composed of~$7$ low-frequency antennas operating in \SI{[1.7-3.7]}{\mega\hertz} and covering approximately \SI{1}{\kilo\metre\squared}. We report the observation of~$18$ air showers detected in coincidence in the three instruments, and estimate a detection threshold of \SI{23\pm4}{\micro\volt\per\metre} from comparisons with detailed SELFAS3 simulations. We also report a strong correlation of the low-frequency signal observation with the atmospheric electric field.
\end{abstract}
}

\begin{document}

\maketitle

\section{Introduction}
\label{Intro}
It is a well known fact that the coherent radio emission during the development of air shower has two main origins: transverse current variation induced by the geomagnetic field, and the charge excess mechanism~\cite{Kahn206,Askaryan,2009APh....31..192A,2015APh....69...50B,radioemissionprd}. The resulting emission appears as a fast electric field transient lasting few tens of nanoseconds, which can be detected by large bandwidth antennas and fast acquisition systems. In the most common way, the observations are carried out in the range \SI{[20-80]}{\mega\hertz} (noted MF in the following, for Medium Frequencies) by experiments such as AERA~\cite{ThePierreAuger:2015rma}, Tunka-Rex \cite{BEZYAZEEKOV201589}, TREND \cite{2012arXiv1204.1599O}, Yakutsk experiment~\cite{Knurenko:2016dck} or LOFAR~\cite{2013AA...556A...2V}. The use of this band is mainly due to man-made broadcasting at low and medium frequencies (AM, FM bands) leading to the choice of relative low sampling rates (\SI{\sim 200}{\mega\sample\per\second}) of the digitizers used by experiments such as AERA and LOFAR. However, CODALEMA~\cite{CODA3} works with a sampling rate of \SI{1}{\giga\sample\per\second}, making it possible to extend observations above the FM band where ARIANNA \cite{BARWICK201750}, ANITA \cite{SCHOORLEMMER201632} and CROME \cite{2014PhRvL.113v1101S} were or are operating. The main limitation of the frequency band is then due to the bandwidth of the antenna used, which is optimized and well mastered in \SI{[20-200]}{\mega\hertz} for CODALEMA, referred to as Extended Medium Frequencies (EMF) in the following.

Several detections at low frequencies (hereafter LF, below \SI{20}{\mega\hertz}) have been carried out in the 70's and up to the 90's. A main conclusion can be drawn from these observations (partially summarized in table~\ref{bfbibtabl}): the results are not well understood. Indeed, several experiments \cite{1970ICRC....3..717P,1970Natur.225..253A,stubbs1971,houghprescottclay1971} have reported that when the frequency decreases, a strong increase of the radio pulse amplitude is observed. For example, Prescott et. al \cite{1970Natur.225..253A} reported transients of about \SI{300}{\micro\volt\per\metre\per\mega\hertz} at \SI{3.6}{\mega\hertz}, averaged over 400 showers whose energy was not known. Nevertheless, the fact that most of those measurements of large electric field have not been reproduced casts doubt on the plausability of such large values \cite{stubbs1971}. Moreover, some estimations of shower energy seem to be incredibly low to permit a radio detection: for example, Stubbs \cite{stubbs1971} reported an energy of a detected primary cosmic ray of supposedly \SI{2\times10^{14}}{\electronvolt}, and a transient amplitude of about \SI{1}{\micro\volt\per\metre\per\mega\hertz} at \SI{2}{\mega\hertz} averaged over 100 showers. However, most of the modern experiments have an energy threshold for detecting a radio contribution from a cosmic-ray-induced air showers around \SI{10^{16}}{\electronvolt} in the MF band. If one considers that the LF signal is, like the MF one, proportional to the number of secondary particles, this would mean that the LF signal has been produced by about at least a hundred times less particles than the smallest detectable MF signal. This would probably imply that either an unknown kind of enhancement of the LF signal acts to make it detectable above the radio noise at these frequencies (see section \ref{Sky}), or another emission process exists (which is discussed further in this article), or more likely, a wrong estimate of the shower energy was made at that time. Assuming erroneous shower energies, the rescaling made in \cite{stubbs1971} to get a normalization for a shower of an energy equal to \SI{10^{17}}{\electronvolt} gives huge and probably overestimated electric field amplitudes of about \SI{500}{\micro\volt\per\metre\per\mega\hertz}. As shown in the current paper, such values would easily be detectable on a reasonable timescale regarding the expected number of showers at \SI{10^{17}}{\electronvolt} falling on any ---~even small~--- detection arrays. This is also in contradiction with the fact that no other LF experiment was able to detect them at that time. Despite these puzzling issues, Akeno experiment \cite{1987ICRC....6..125N} found correlation of signals in the LF band with signals at higher frequencies. They reported signal amplitudes of about~\SI{100}{\micro\volt\per\metre\per\mega\hertz} in the range~\SI{[26-300]}{\kilo\hertz}, a measurement that has been successfully repeated and which is consistent with other measurements carried out by EAS-RADIO \cite{castagnoli1991}.\\

\begin{table}[]
\centering
\begin{tabularx}{\textwidth}{|c|c|X|}
\hline
Ref & Frequency & Observations \\ \hline
\cite{1970ICRC....3..717P}    &     \SI{3.6}{\mega\hertz}      &      Calgary (\SI{1049}{\metre} asl), signal at \SI{3.6}{\mega\hertz} 3 times larger than at \SI{22}{\mega\hertz},            \\ 
                              &                                &      no signal larger than noise at \SI{10}{\mega\hertz}.     \\ \hline
\cite{1970Natur.225..253A}    &     \SI{2}{\mega\hertz}      &      Haverah Park: signal at \SI{2}{\mega\hertz} 200 times larger than at \SI{32}{\mega\hertz} \\
                              &                                &      and 375 times larger than at \SI{44}{\mega\hertz}       \\ \hline
\cite{stubbs1971}             &     \SI{2}{\mega\hertz}      &      Buckland Park: \SI{1}{\micro\volt\per\metre\per\mega\hertz}, estimated energy of \SI{2\times10^{14}}{\electronvolt},   \\ 
                              &                                &      giving after rescaling \SI{500}{\micro\volt\per\metre\per\mega\hertz} at \SI{10^{17}}{\electronvolt}.\\
                              &				     &	     From that, signal at \SI{2}{\mega\hertz} would be 250 times larger than at \SI{44}{\mega\hertz}      \\ \hline
\cite{houghprescottclay1971}  &     \SI{3.6}{\mega\hertz}      &     Dominion Radio Astro Observatory: signal at \SI{3.6}{\mega\hertz} 1 order of magnitude          \\
                              &                                &       larger than in \SI{[20-60]}{\mega\hertz}, but less by a factor of 3-4 than at \SI{2}{\mega\hertz}          \\ \hline
\cite{1987ICRC....6..125N}    &     \SI{[0.026-0.3]}{\mega\hertz}      &      Akeno, AGASA: unipolar and negative signal, with a width of \SI{5}{\micro\second},          \\
                              &                                &      field decreasing with decreasing distance   \\ \hline
\cite{castagnoli1991}         &     \SI{2.6}{\mega\hertz}      &       EASRADIO: increase in electric field strength when frequency decreases.     \\
                              &                                &      Observation of low-frequency signal in coincidence with extensive air shower.               \\ \hline
\end{tabularx}
\caption{Summary of some pioneer observations in the LF band: references, observations frequency (\si{\mega\hertz}), comments.}
\label{bfbibtabl}
\end{table}

\begin{figure}[!ht]
\begin{center}
  \includegraphics[width=0.7\textwidth]{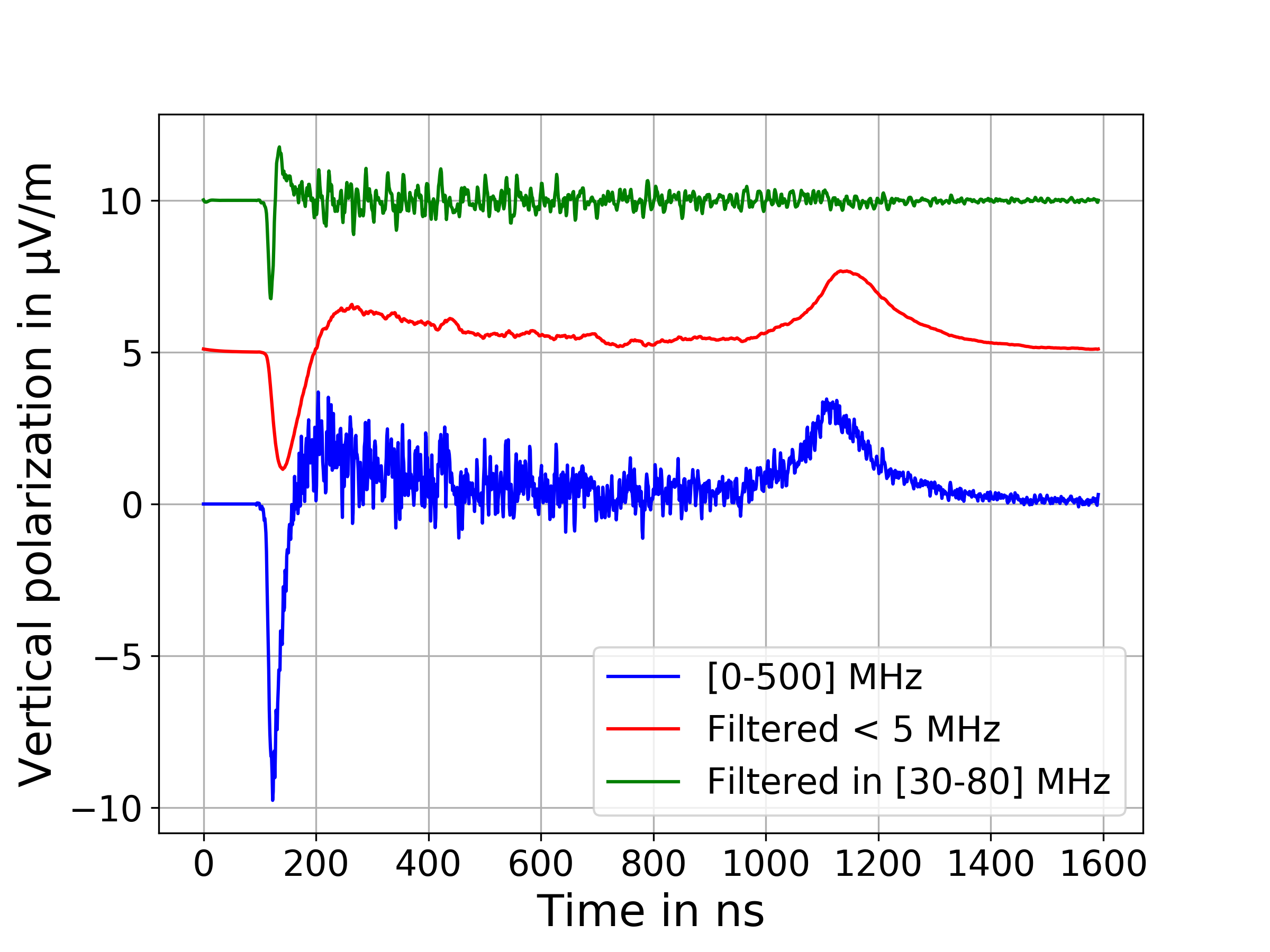}
\caption[]{Vertical polarization of the electric field as a function of time  obtained with a SELFAS3 simulation for an observer at \SI{300}{\meter} of the shower core and for a vertical proton-induced shower at \SI{10^{18}}{\electronvolt} (blue) for the location of the Nan\c{c}ay Radioastronomy Observatory. Associated filtered responses in different bands (green for \SI{[30-80]}{\mega\hertz} and red for $<$\SI{5}{\mega\hertz}) show that two pulses can be seen below \SI{5}{\mega\hertz}. The red and green curves are positively shifted along $y$ for clarity.}
\label{simu}
\end{center}
\end{figure}

From these observations, and in spite of the controversy, we have therefore chosen to retake the LF study, taking advantage of an improved version of SELFAS3 simulation code of the radio signal of cosmic-ray-induced air showers \cite{Marin2012733,Garcia-Fernandez:2017yss}, and by installing LF antennas in the CODALEMA experiment \cite{CODA3,AntonyICRC2017}. Fig.~\ref{simu} shows the vertical polarization of the electric field obtained with SELFAS3 for an observer at \SI{300}{\meter} of the shower core and a vertical proton as primary cosmic ray at \SI{10^{18}}{\electronvolt} for the location of the Nan\c{c}ay Radioastronomy Observatory\footnote{Altitude:~\SI{130}{\metre}~asl, geomagnetic field amplitude of $24~\mu$G with a unit vector oriented as: $(B_x=0.0030,B_y=0.4548,B_z=-0.8906)$, $x$ being the east-west direction (positive towards east), $y$ the geographical north-south direction (positive towards north) and $z$ the local vertical).}.

On the full-band trace (blue line), the negative peak at \SI{150}{\nano\second} is due to the shower development in air. The second, positive peak around \SI{1150}{\nano\second} has been interpreted as a new mechanism: the effect of the coherent deceleration of the shower front when hitting the ground, called the ``sudden death pulse'' (SDP) \cite{2012arXiv1211.3305R,Garcia-Fernandez:2017yss}. In our antennas located at 300~m from the core, the simulated SDP arrives $\delta t=300/c=\SI{1}{\micro\second}$ after the normal pulse, which is consistent with the propagation time from the shower core. After filtering (Butterworth 1st order) in different frequency bands, only the shower development peak survives in the MF band while both signals are still detectable for frequencies below $<$ \SI{5}{\mega\hertz}, confirming the interest of using LF antennas. Detecting and studying this phenomenon on one hand and understanding the low frequency counterpart of the radio emission of air showers on the other hand are the goals of the EXTASIS (EXTinction of Air-Shower Induced Signal) experiment.

The LF band presents two interests. The first one is the SDP predicted by the simulation. The sudden coherent deceleration of the secondary electrons in the shower front when reaching the ground level emits a strong electric field which could be detectable. Let the reference time ($t=0$) be the instant of impact at ground. The SDP arrives at the antenna at time~$t_{\mathrm{SDP}}\backsimeq d_{\mathrm{core}}/c$, with~$d_{\mathrm{core}}$ the distance between shower core and the antenna and $c$ the speed of light in the medium. If at least 3 antennas are involved, the core position can be estimated by simple intersection of circles of radius~$d_{\mathrm{core}}$. If we can observe with the same antenna the electric field from the development in the air and the SDP, we could have an intrinsic time scale within the shower, which would allow a direct estimation of the distance between the~$X_{\mathrm{max}}$ (the atmospheric depth of the maximum of the shower development) and the ground along the shower axis~\cite{2012arXiv1211.3305R}, providing an excellent way to obtain the nature of the primary cosmic ray. While the electric field amplitude on the ground created by the regular emission needs a two-dimensional description in general, the amplitude of the SDP decreases as~$1/d_{\mathrm{core}}$. The MF pulse amplitude strongly depends on the axis distance~$d_{\mathrm{axis}}$ with a Gaussian decrease at first order, and simulation also predicts that the regular pulse can be detected at larger distances at low frequencies, as it will be shown in the following, where we detail the objectives of the EXTASIS experiment, our instrumental setup and our first results.

\section{General characteristics of the sky at LF}
\label{Sky}
Being extensively used, the MF and EMF bands are nowadays well known in terms of background noise. This is not the case of the LF band and the studies made in the 70's probably need to be reevaluated considering the strong evolution of the radiocommunications over the last 50 years. Consequently, the first work to carry out is to explore the LF environment, in particular the LF sky of our experiment site. The frequency range studied is dominated by the high brightness temperature of the atmospheric noise. In this section, we remind some facts on the ionosphere layers, and then focus on the background noise at LF. 

\subsection{Absorption of radio emission lines at low frequencies in the ionosphere}
\label{pwlfr}
The overall background radio noise at low frequencies is less important during the day than during the night. This can be explained by the absorption of radio emission lines at these frequencies by the ionosphere. Indeed, as explained in \cite{Seybold}, the ionosphere is composed of layers of ionized plasma constrained by the geomagnetic field. The properties of the layers depend on the free electron density, the altitude and the season. Notably, during the night, some layers disappear or shrink, and some others combine. These changes lead to the absorption, refraction, attenuation, depolarization and dispersion of radio emission lines. 
For example, the D layer of the ionosphere (\SI{[70-90]}{\kilo\metre} of altitude) absorbs the radio emission lines from~\SI{0.3}{\mega\hertz} to~\SI{4}{\mega\hertz}, which tends to decrease the noise level in this frequency band. However, the D layer is only present during the day and vanishes or diminishes at night, no longer absorbing the radio emission lines. The behaviour of the D layer is only partly responsible for the daily variation of the noise. The F layer, which is composed of two sub-layers combining at night (from 150 to \SI{\sim470}{\kilo\metre} of altitude), will tend to increase the sky-wave propagation during the night by reflecting and refracting them. Consequently the noise level after dark increases, due to both the reflection/refraction by the F layer and the disappearance of the D layer. This daily variation is expected to be visible in LF antenna data.

\subsection{Overall background noise}
\label{bkgan}

Between few hundred of \SI{\kilo\hertz} and \SI{10}{\mega\hertz}, the background noise is the superposition of three dominant contributions:
\begin{itemize}
\item{the minimal thermal noise of the atmosphere (in other words, its brightness related to its temperature),}
\item{the galactic radio emission,}
\item{the so-called atmospheric noise, made of the contribution of the electric discharges from thunderstorms at planetary scale~---~$\sim 100$ per second~---, and of any kind or man-made noise (radio transmitters, anthropic transients).}
\end{itemize}
Thunderstorm discharges and man-made noise are not absorbed by the D atmospheric layer at night, and are reflected and refracted between the ionosphere layers and the ground. This causes an increase of the atmospheric noise during night. Contrarily to the galactic noise, the atmospheric background noise is not predictable and varies as function of the location on Earth, the season and the time of the day (see also section~\ref{psdsky}).

\begin{figure}[!ht]
\begin{center}
  \includegraphics[width=\textwidth]{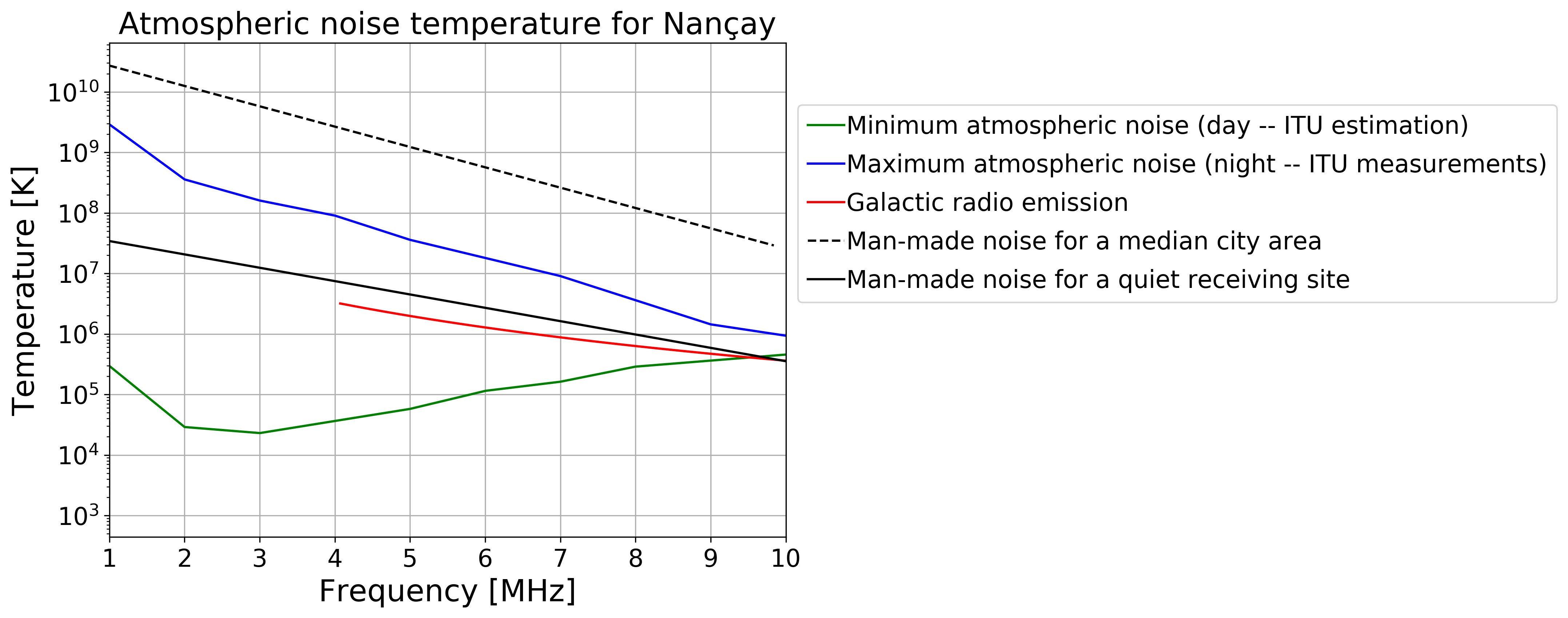}
\caption{Atmospheric noise and galactic noise temperature as a function of the frequency, computed from the raw ITU data and corrected for the site of Nan\c cay. Man-made noise temperature estimates have been added for a city and a quiet site. The galactic emission (considered apart from the atmosphere) is computed with the model established by Cane \cite{1979MNRAS.189..465C}, on the basis of ground measurements above \SI{4}{\mega\hertz} and satellite measurements below this frequency, where the atmosphere becomes mostly opaque to the incoming radiation. The galactic noise contribution thus experiences a cut-off below \SI{4}{\mega\hertz} for realistic observation conditions, and is no longer dominant regarding to the atmospheric noise and man-made noise for a quiet receiving site.}
\label{sky}
\end{center}
\end{figure}

Fig.~\ref{sky} presents the background noise brightness temperature as a function of the frequency. Data are extracted from the International Telecommunication Union (ITU) \cite{ITU} and corrected for the site of Nan\c cay. For the atmospheric noise, only the contribution of the electric discharges from thunderstorms is taken into account in this reference, though an estimate of the made-made noise is given for several types of sites.

If there were absolutely no other sources of noise, the absolute limit for a detectable signal would be set by the minimal thermal noise density $N_{\mathrm{ref}}$, defined as
\begin{equation}
N_{\mathrm{ref}}=10\log\left(\frac{kT_0}{1\;\mathrm{mW}}\right)=\SI{-174~}{\deci\bel m\cdot\hertz^{-1}}
\end{equation}
where $T_0=$\SI{290}{\kelvin} is the reference air temperature for a reference power of \SI{1}{\milli\watt} and $k$ is the Boltzmann constant. In all cases this limit is surpassed by the galactic background, until about \SI{150}{\mega\hertz}. Let $F_{\mathrm{am}}$ ("am" stands for atmospheric and median) be the difference between the noise coming from thunderstorms and the minimal thermal noise $N_{\mathrm{ref}}$. $F_{\mathrm{am}}$ is then expressed as
\begin{equation}
F_{\mathrm{am}}=10\log\left(\frac{T_\mathrm{n}}{T_0}\right)\quad \mathrm{dB}
 \end{equation}
where $T_\mathrm{n}$ is the temperature of the sky. $F_{\mathrm{am}}$ depends on this temperature, on the location of the observing site, on the day-night cycle and on the seasons. At \SI{1}{\mega\hertz}, the daily variation of $F_{\mathrm{am}}$ is \SI{40}{\deci\bel} for winter and \SI{35}{\deci\bel} for summer: this high variability of the atmospheric noise is easily observable with LF antennas. At frequencies above \SI{20}{\mega\hertz} (MF), the observation limit is set by the galactic background noise, whose variations are visible by the CODALEMA antennas for instance: the atmospheric noise is no more dominant whatever the hour of the day. At \SI{4}{\mega\hertz}, the brightness temperature of the atmospheric noise is \SI{20}{\deci\bel} below the galactic noise level in the best daytime case while it is \SI{17}{\deci\bel} above during the night. These are the real detection limits at LF which are anyway surpassed by the man-made noise level even for a quiet receiving site.
To sum up, we show in Fig.~\ref{sky} that for a quiet observation site, the noise limit is set by man-made activities during the day (the galactic noise being barely competitive), while during the night it is dominated by the atmospheric noise, and in all cases the LF noise level is well above the noise at MF.
We can assess that the detection of air-shower transients will be complicated at LF, especially if their strength does not increase when the frequency drecreases, as it will be shown in the following.

\subsection{Power spectrum density of the low frequency sky at Nan\c cay}
\label{psdsky}
\begin{figure}[!ht]
\begin{center}
\subfloat[LQ antenna, summer]{
  \includegraphics[width=0.5\textwidth]{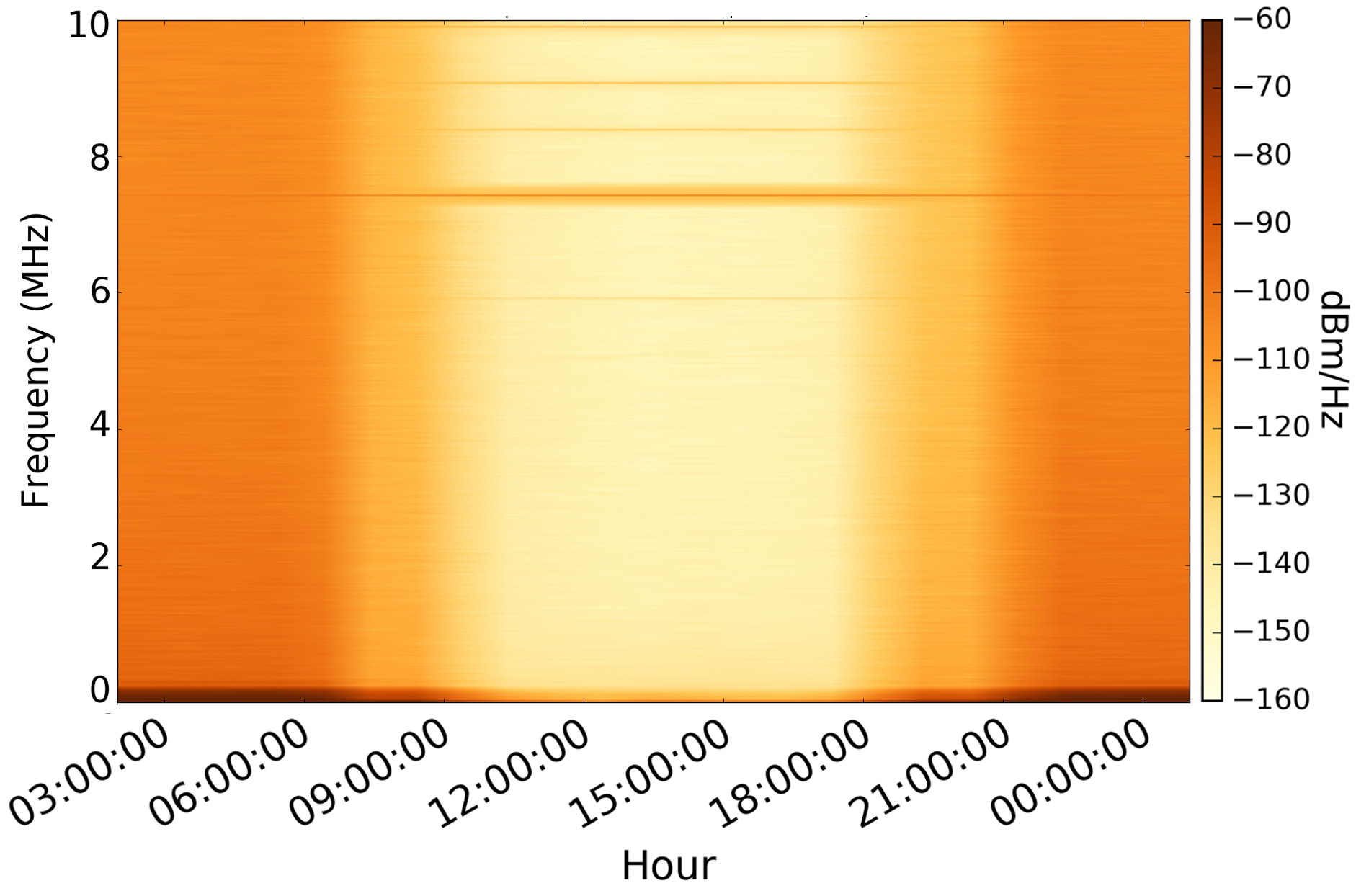}
}
\subfloat[QH antenna, summer]{
  \includegraphics[width=0.5\textwidth]{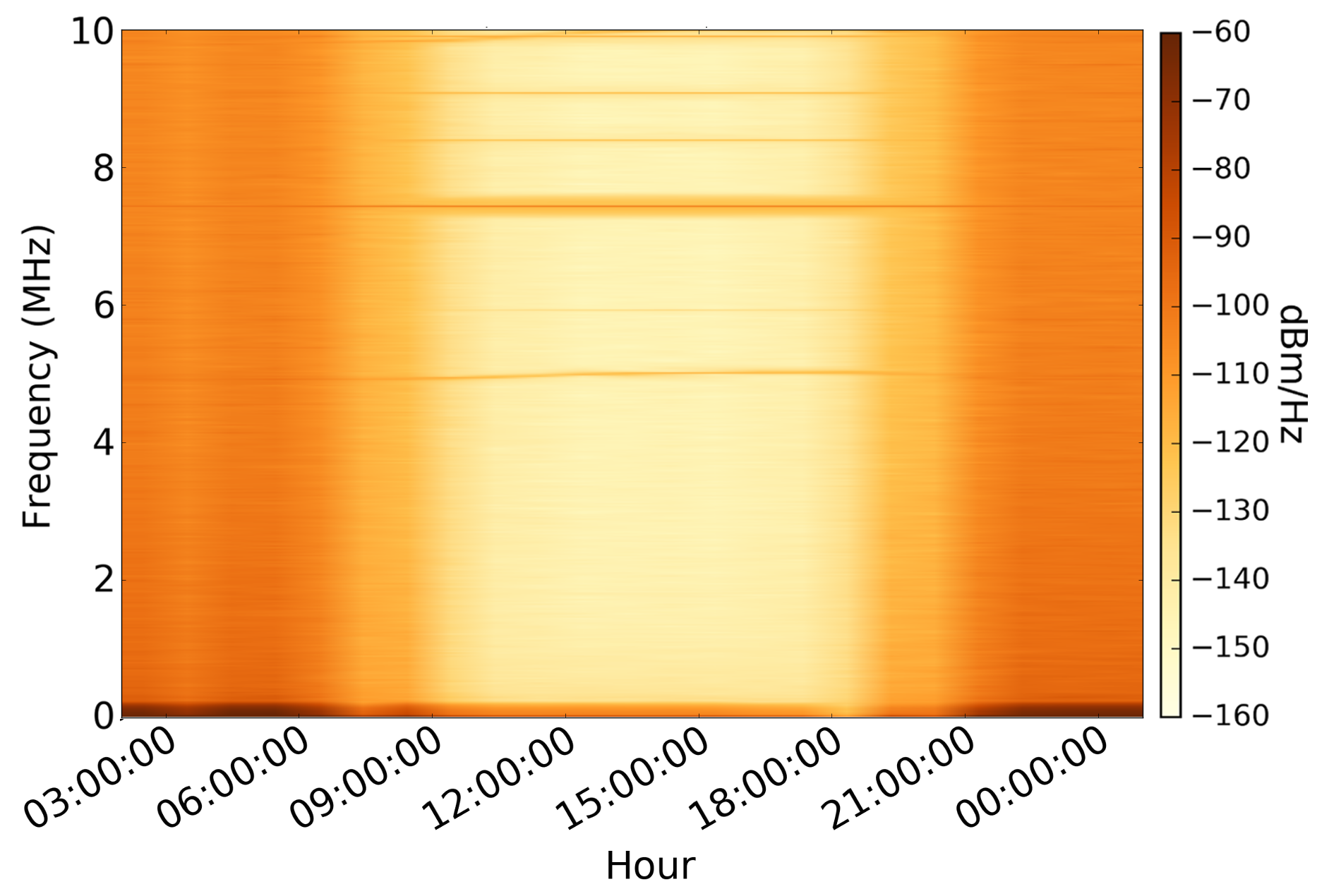}
}
\\
\subfloat[HL antenna, summer]{
  \includegraphics[width=0.5\textwidth]{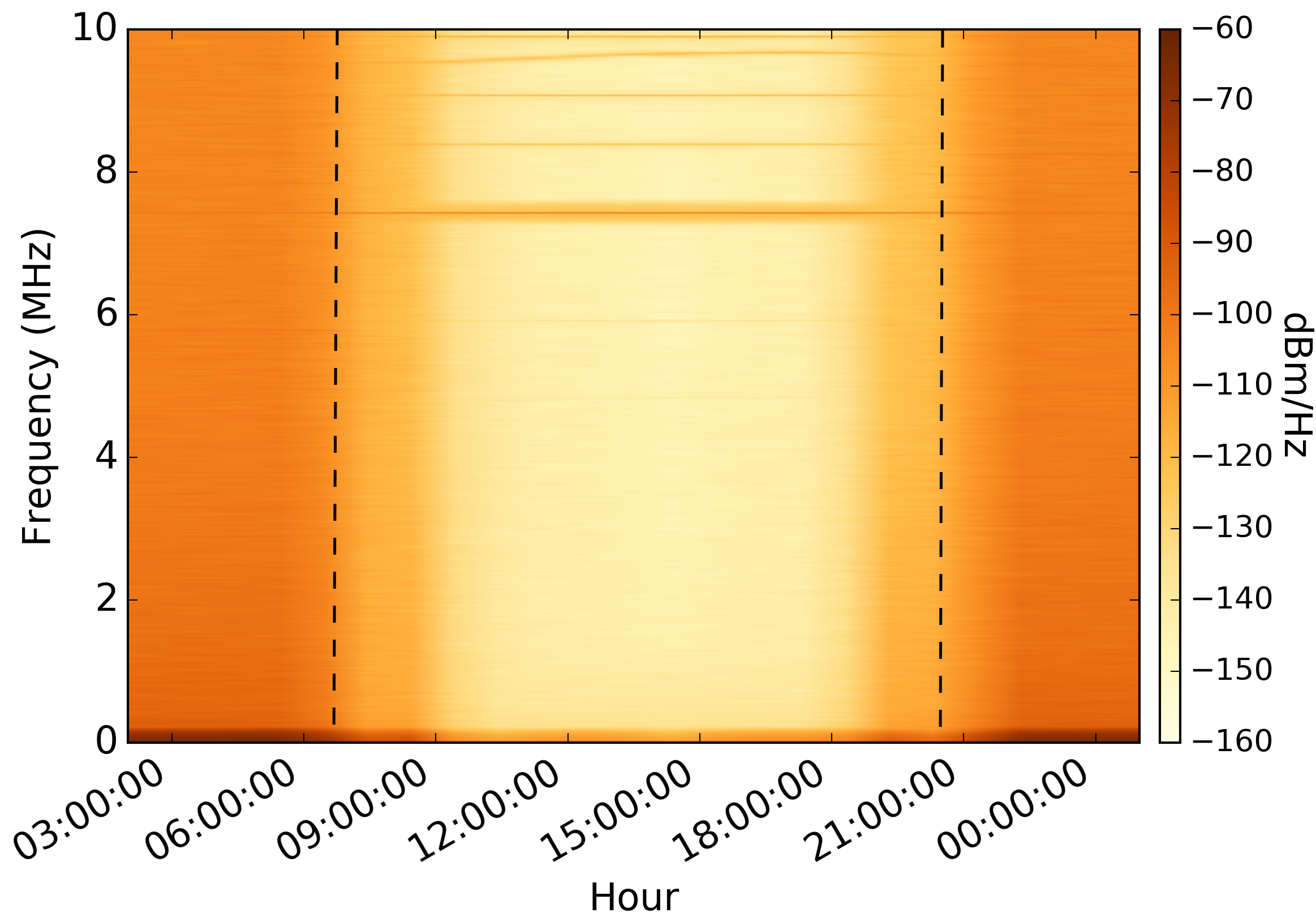}
}
\subfloat[HL antenna, winter]{
  \includegraphics[width=0.5\textwidth]{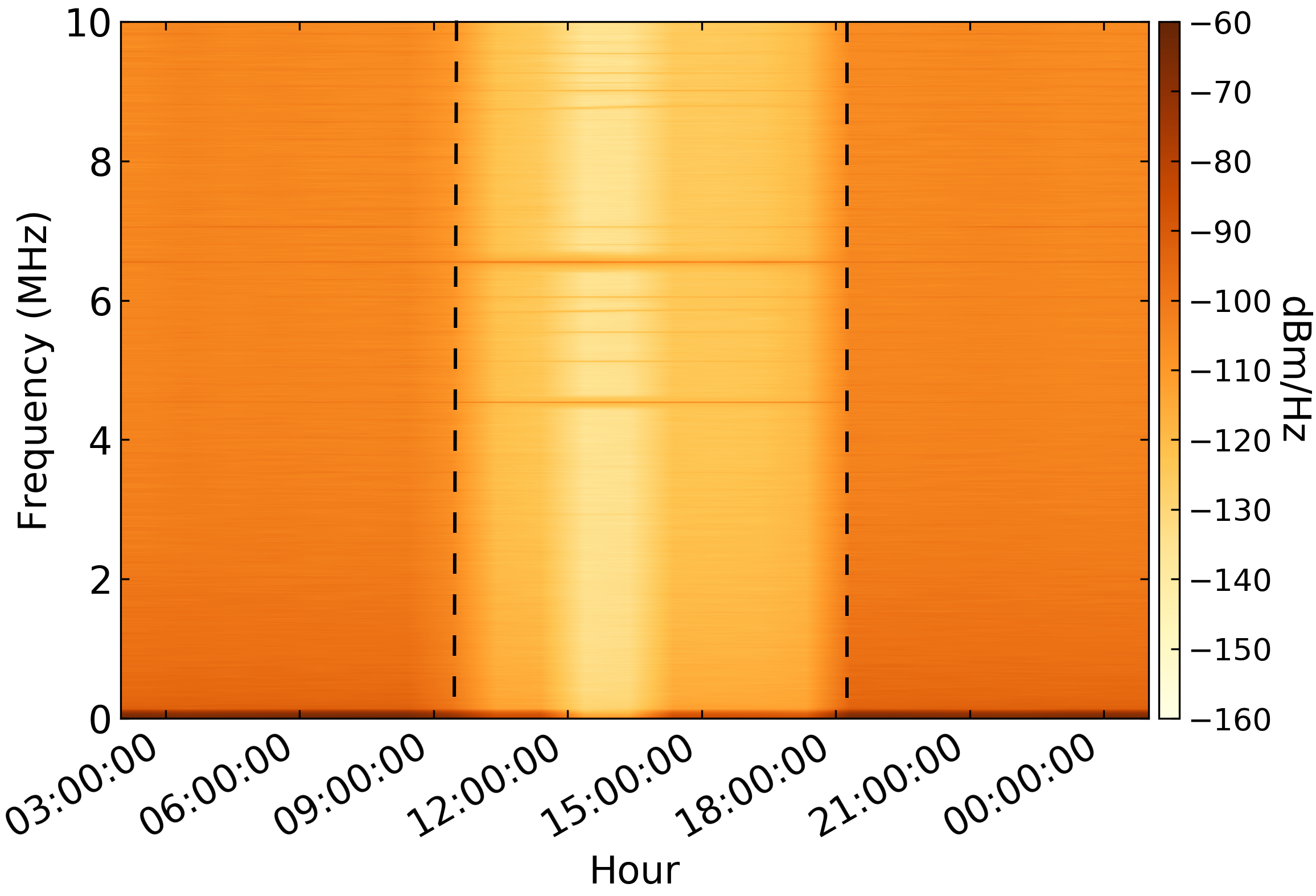}
}
\caption[Time-frequency diagram for different LF antennas]{Time-frequency diagram for different LF antennas. Fig. (a), (b) and (c) report environmental variations due to the location of the antennas, while Fig. (c) and (d) report a seasonal variation for the same antenna. The vertical black-dashed lines represent the sunrise and sunset. See text for details.}
\label{LFsky}
\end{center}
\end{figure}

Fig.~\ref{LFsky} presents time-frequency diagrams for different LF antennas of EXTASIS. The time-frequency diagram consists in a daytime power spectrum density (PSD, in \si{\deci\bel m\cdot\hertz^{-1}}). The darker the color, the noisier the environment. The diagram gives a view of the environment of each antenna, which varies with the position of the LF antenna (pictures (a) to (c)) and during one day, regardless of the position of the antenna (day/night dependence is well observed). Moreover, a seasonal variation can also be seen, as shown in pictures (c) and (d), which presents time/frequency diagrams for one LF antenna taken during summer (c) and winter (d). The vertical black-dashed lines represent the sunrise and sunset times, delimiting night time and daytime. Of course, the two time periods do not have the same duration depending on the season, the duty cycle being maximal during summer and minimal during winter. 

\begin{figure}[!ht]
\begin{center}
\subfloat[Nighttime PSD.]{
  \includegraphics[width=0.497\textwidth]{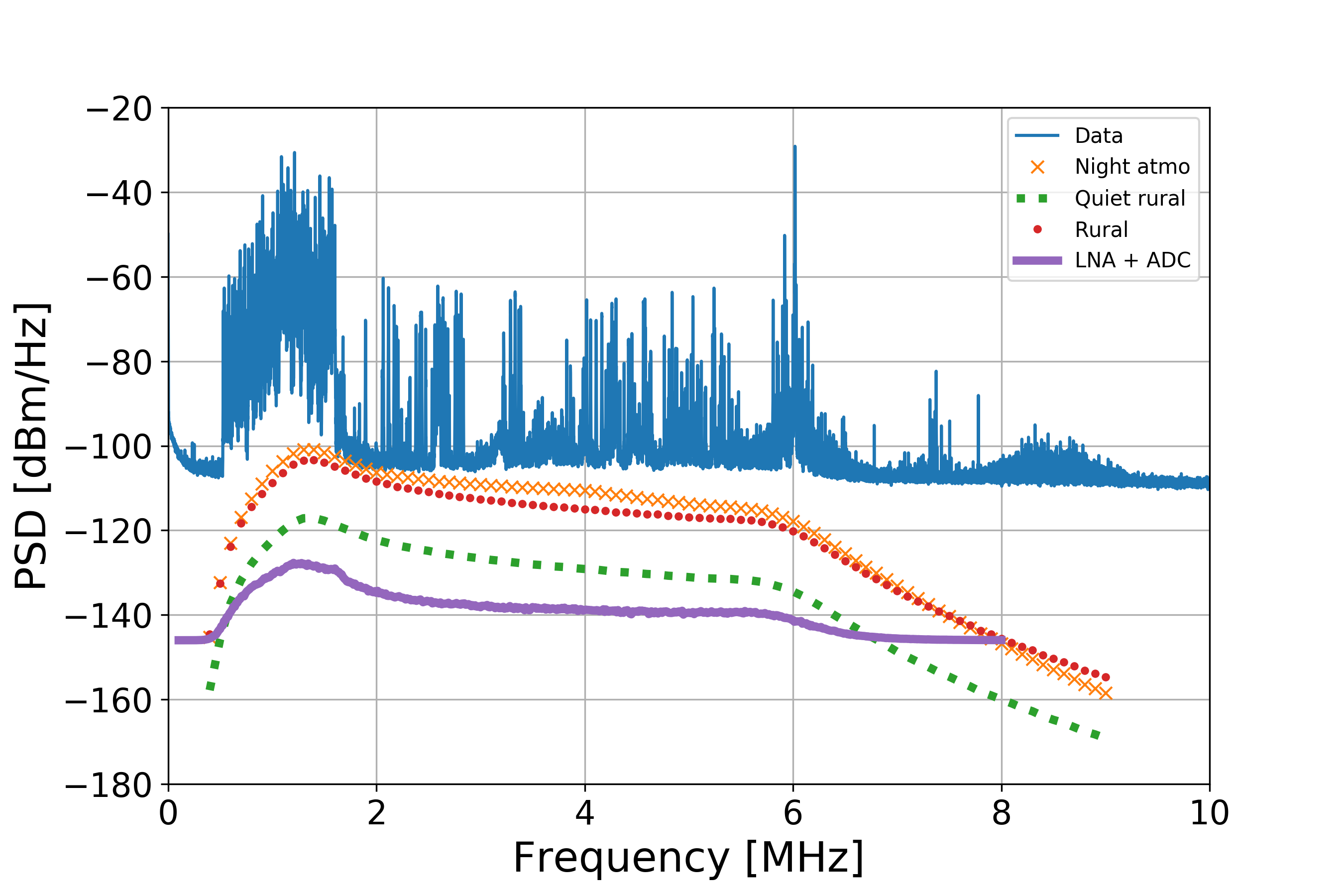}
}
\subfloat[Daytime PSD.]{
  \includegraphics[width=0.5\textwidth]{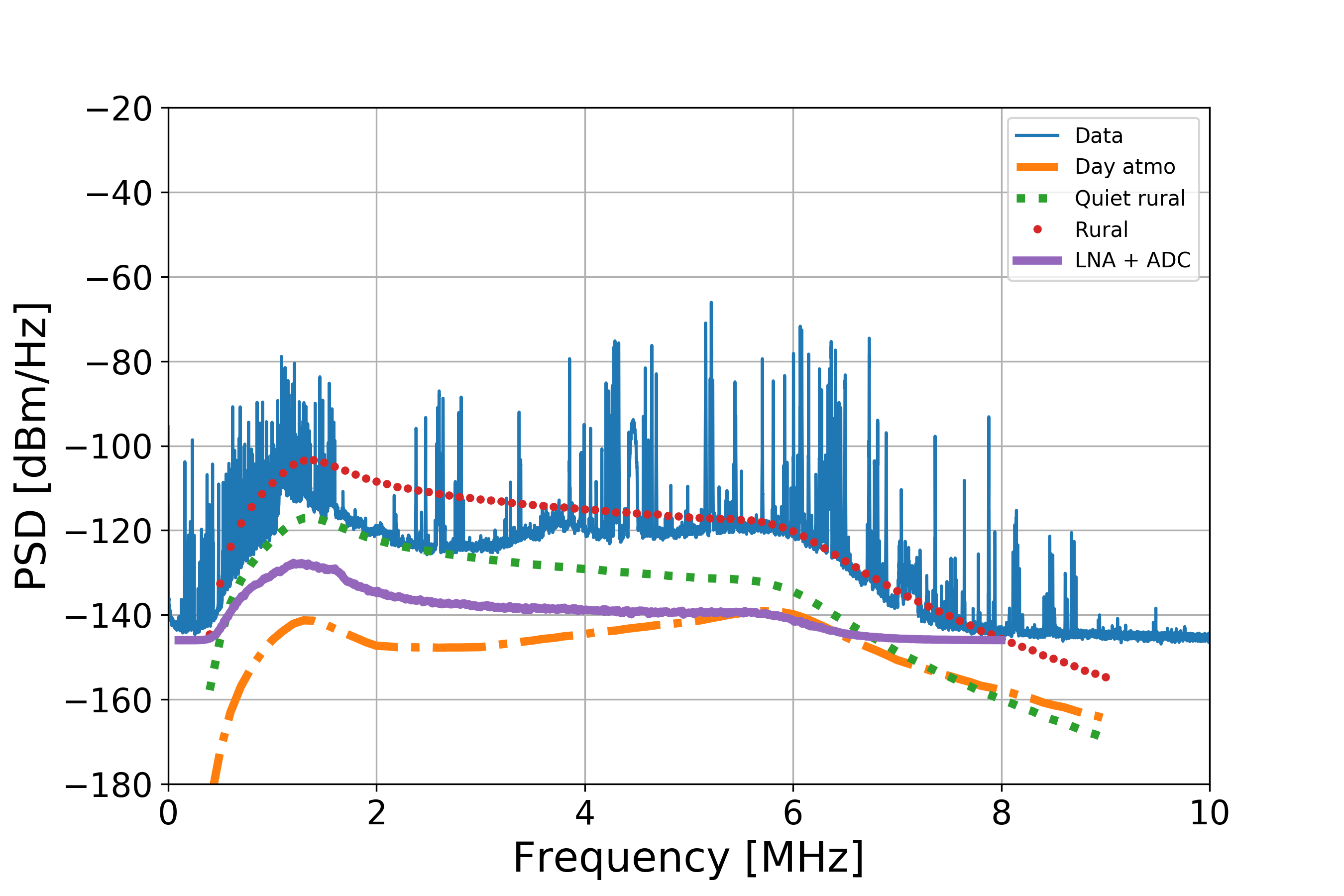}
}
\caption[]{Night (left) and day (right) PSD of one LF antenna in vertical polarization, averaged over 200 events, for the same day than Fig.~\ref{LFsky}.(a). The atmospheric noise of Fig.~\ref{sky} is shown together with the noise from the electronic chain. The figure on the right shows that the Nan\c{c}ay site can be considered as a quiet rural site up to 3.2~MHz.}
\label{LFspectre}
\end{center}
\end{figure}

From Fig.~\ref{sky}, one could wonder whether the Nan\c cay site is classified as a quiet (rural) or urban site, which would drastically hamper any observation at LF. Fig.~\ref{LFspectre} presents the PSD of one LF antenna as recorded by our analog and digitization chain, averaged over 200 events during nighttime (a) and daytime (b), for the same day and same antenna than in Fig.~\ref{LFsky}.(a). The atmospheric and estimated man-made noises of Fig.~\ref{sky} (calculated at the ADC input from the antenna simulations and the measured transfer function of the analog chain) are plotted along with the noise from the electronic chain. A lot of man-made broadcasting radio emissions (RFI) are present in the LF band, day and night. However, during daytime, the quiet rural noise level expected from ITU specifications is reached in most of the \SI{[1.7-3.7]}{\mega\hertz} band, which contains less RFI and presents a noise floor \SI{20}{\deci\bel m\cdot\hertz^{-1}} below the noise floor during nighttime. This is the reason why we have chosen to restrict our observations to this band. From these daytime PSDs, one can conclude that the Nan\c cay site can be considered as a quiet rural site regarding the specifications of ITU. We also observe that the LNA noise is at least 15~dB lower than the minimum noise, which is not the case for the horizontal polarization (not shown here) for which the limitation is given by the LNA noise from 1.5 to 2.1~MHz. Therefore for the vertical polarization, the minimum signal to be detected is not limited by the sensitivity of our detector. It is worth noticing that, in \SI{[1.7-3.7]}{\mega\hertz}, the root mean square (rms) of the noise at night is $\sim100$ times higher than the noise during the day. It means that, to be detected during the night, a pulse should be 100 times higher than a pulse observed during the day. In fact, as it will be shown in section~\ref{lowratediscussion}, no valuable observations have been made during night time.

\section{Instrumental setup}
\label{Setup}

\begin{figure}[!ht]
\begin{center}
\subfloat{
  \includegraphics[width=0.625\textwidth]{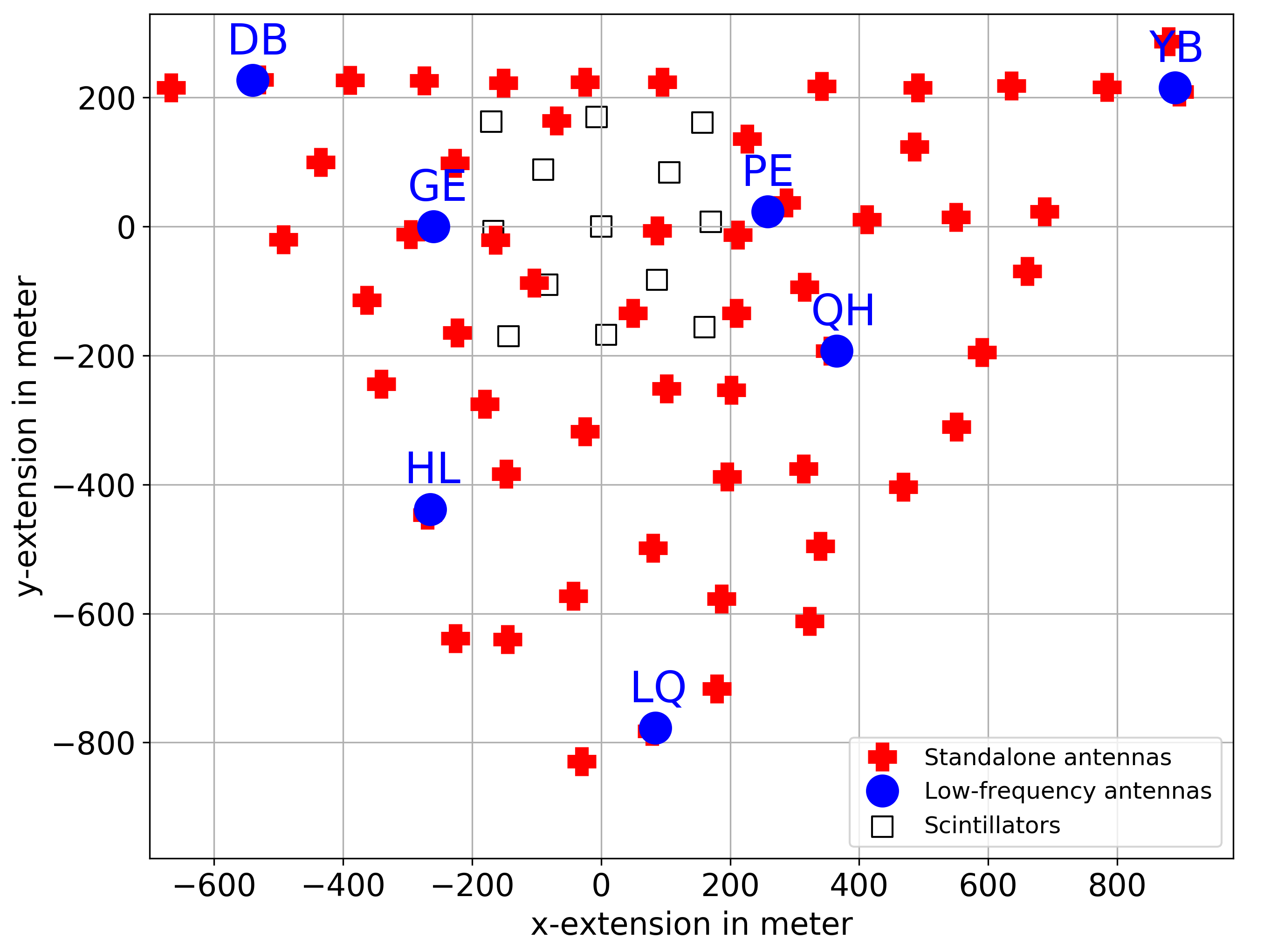}
}
\subfloat{
  \includegraphics[width=0.35\textwidth]{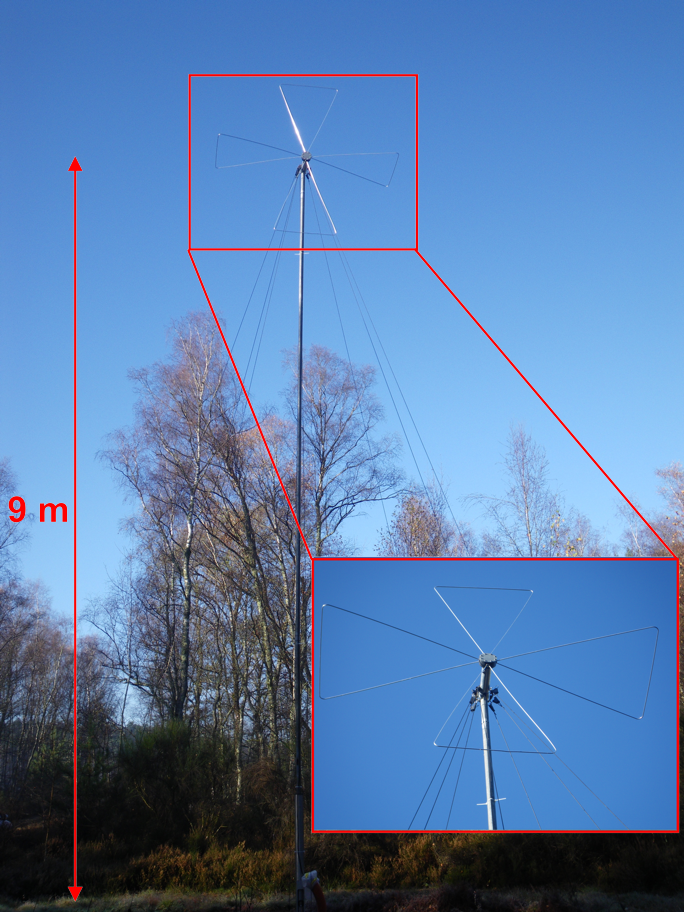}
}
\caption[Area experiment at Nan\c cay and picture of a LF antenna]{Left: experimental area at Nan\c cay. Red crosses represent the 57 standalone antennas, black squares represent the 13 scintillators, the blue points represent the 7 LF antennas. Right: photography of a LF antenna.}
\label{nancay2}
\end{center}
\end{figure}

Hosted since 2002 by the Nan\c{c}ay Radioastronomy Observatory, CODALEMA is one of the pioneering experiments that have participated in the rebirth of radio detection of cosmic rays at the beginning of the 21$^\text{st}$ century. Over the years, the experiment has seen the development of a large collection of detectors, intended to study the properties of the radio emission associated with cosmic ray induced air showers in the energy range from $10^{16}$ to \SI{10^{18}}{\electronvolt}. In its current version, CODALEMA consists essentially of:
\begin{itemize}
\item{a square array (0.4 $\times$ \SI{0.4}{\kilo\metre\squared}) of 13 particle scintillator counters ;}
\item{a set of 57 so-called ``standalone'' cross-polarized antennas, operating in the EMF band, distributed over \SI{1}{\kilo\metre\squared} ; }
\item{a so-called ``Compact Array'' of 10 cross-polarized antennas, arranged in a star shape of \SI{150}{\metre} extension and whose signal acquisition (in the MF band) is triggered by the particle detector.}
\end{itemize}
Details on these arrays are given elsewhere \cite{CODA3}. CODALEMA is today the supporting experiment of the EXTASIS experiment, an array of 7 low-frequency antennas triggered by the particle detector, which takes advantage of its existing infrastructure. The LF antenna locations have been chosen to cover the overall Nançay area and also in a way that each LF antenna has a MF standalone antenna nearby. LF antennas are named DB, YB, GE, PE, HL, QH and LQ. Fig.~\ref{nancay2} shows the experimental area at Nan\c cay (the compact array is not represented).

\subsection{The LF antenna}
\label{LFantenna}
\begin{figure}[!ht]
\begin{center}
  \includegraphics[width=1\textwidth, angle=0]{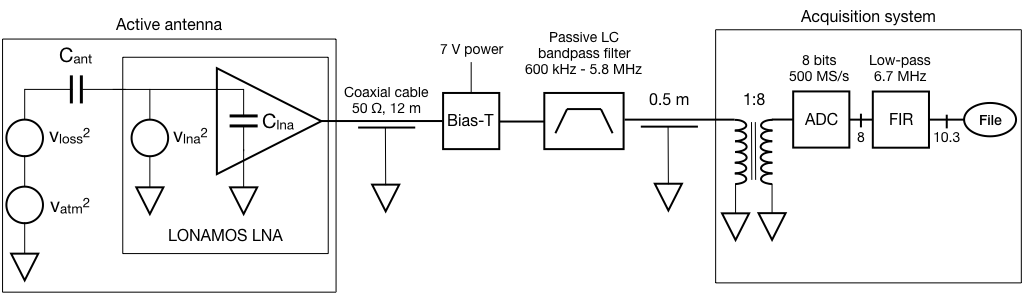}
\caption[sketch of the overall system]{Simplified sketch of the active antenna ant its noise sources, the RF components, the ADC and signal processing. $v_\text{atm}^2=4k\,T_\text{atm}\,R_\text{rad}$ is the equivalent noise source of the overall atmospheric noise seen in 2$\pi$~sr by the antenna. $v_\text{loss}^2=4k\,T_{0}\,R_\text{loss}$ is the equivalent noise source of the ground losses seen by the antenna. $v_\text{lna}^2$ is the equivalent noise source of the noise of the LNA located at the feedpoint of the antenna. $v_\text{atm}^2$, $v_\text{loss}^2$ and $v_\text{lna}^2$ are noise densities in \si{\volt\squared\per\hertz} unit. Triangles pointing down feature grounding. See text for details on FIR and acquisition.}   
\label{sketch}
\end{center}
\end{figure}
Currently, the EXTASIS experiment is made of dedicated LF antennas (Fig.~\ref{nancay2}) triggered by scintillators. Their design is based on the so-called ``Butterfly'' active antennas in use in CODALEMA (see the concept of these active antennas in~\cite{CharrierARENA2010}), with the same radiating element and same ``LONAMOS'' Low Noise Amplifier (LNA), except that the setting of the LNA is tuned for the frequency band below 10~MHz. The shape of the radiating element is a bow tie made of an aluminium rod with an overall length from end to end of 1.2~m. Apart the LNA, another difference with the CODALEMA antennas is that their dual, crossed polarizations are East-West and Vertical, since the SDP is expected to be mainly vertically polarized \cite{Garcia-Fernandez:2017yss}. Regarding the LF band, this antenna is a short dipole, since its length is less than one tenth of the shortest wavelength. In that case, our antenna impedance is well described by a pure capacitance $C_\text{ant}$ of 12~pF, value estimated using the complex impedance produced by NEC-4 simulations. As the antenna is active, the LNA is located at the antenna feedpoint, as illustrated on the left part of Fig.~\ref{sketch}. Thanks to a low capacitive input impedance of the LNA ($C_\text{lna}=6.6$~pF), the signal received by the antenna is transferred to the LNA input through a capacitive divider giving broadband characteristics. On the sketch of Fig.~\ref{sketch}, the LNA noise density is modelized only by a voltage source $v_\text{lna}^2$, which is a good approximation as the input transistor of the LONAMOS is a CMOS one. Considering the atmospheric noise as a signal, the signal to noise ratio (SNR) of the active antenna can thus be written as:

\begin{equation}\label{sig2noiseequ} 
\frac{v_\text{atm}^2}{v_\text{noise}^2}=\frac{T_\text{atm}\,R_\text{rad}}{T_{0}\,R_\text{loss}+\left(\frac{C_\text{ant}+C_\text{lna}}{C_\text{ant}}\right)^2\frac{v_\text{lna}^2}{4k}}
\end{equation}

$T_\text{atm}$ is the minimum atmospheric day temperature of Fig.~\ref{sky}, $T_0$~=~\SI{290}{\kelvin} the air temperature as previously defined, $R_\text{rad}$ the antenna radiation resistance and $R_\text{loss}$ the loss resistance due to the ground below the antenna. From Eq.~\ref{sig2noiseequ}, the LNA intrinsic noise increases independently of the frequency by a constant factor of 2.4 depending only on the capacitive divider ratio. That would be impossible to obtain with a 50 or 75~$\Omega$ input impedance LNA, unless using an antenna near its first resonance, which would imply a huge and heavy antenna with an end to end length around 60~m for the [1-6]~MHz band. But a drawback of this short dipole is the low value of $R_\text{rad}$, around 1~$\Omega$, as the antenna is used at a frequency~18~times lower than its 45~MHz resonance frequency. Hopefully, this low value of $R_\text{rad}$ is compensated by the huge value of $T_\text{atm}$, keeping the product $T_\text{atm}\,R_\text{rad}$ of Eq.~\ref{sig2noiseequ} high enough against $v_\text{noise}^2$, making it possible to use such a short dipole from the SNR point of view.

\begin{figure}[!ht]
\begin{center}
  \includegraphics[width=0.5\textwidth]{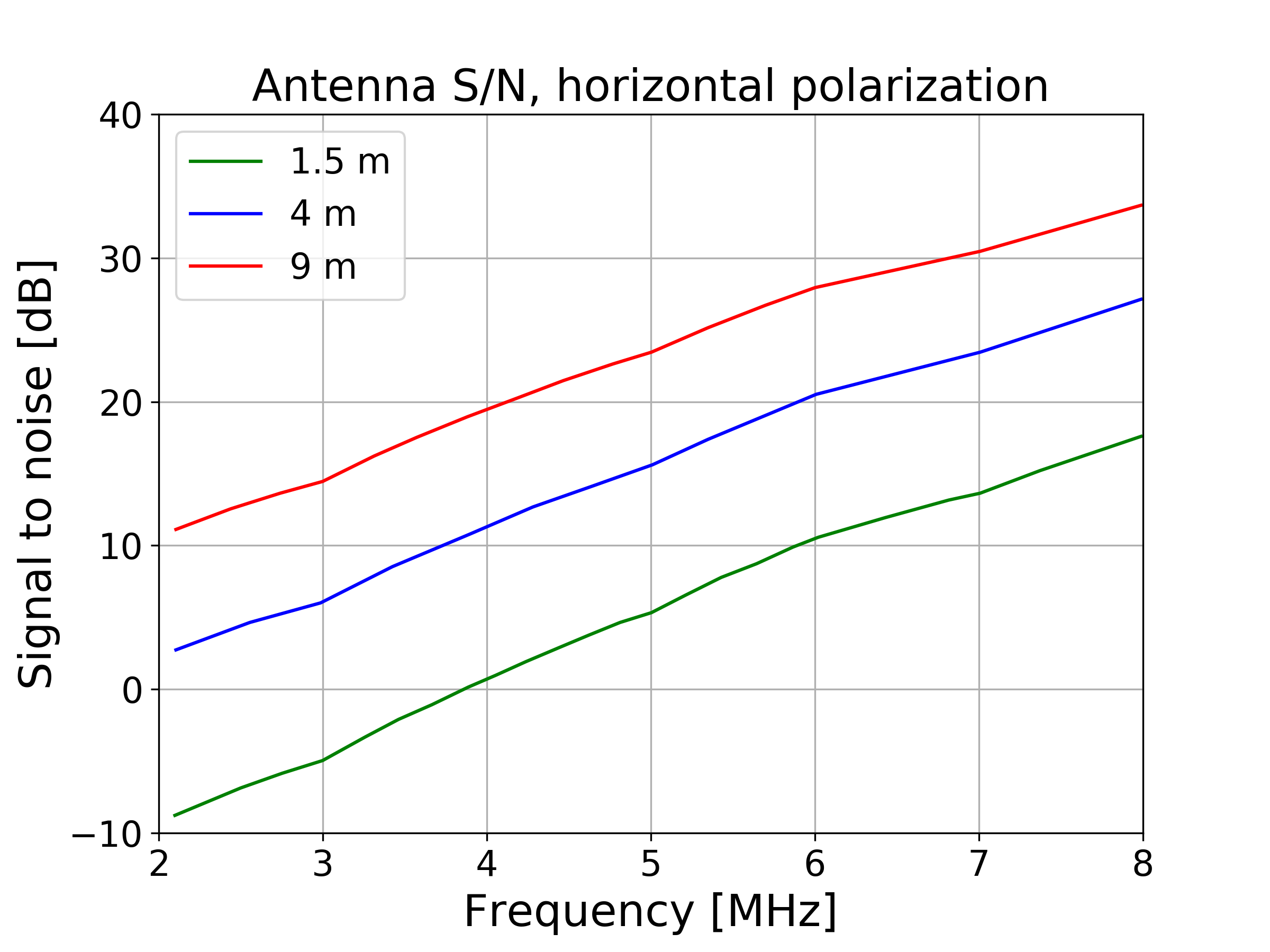}
\caption[Signal to noise ratio of a Butterfly antenna]{Signal to noise ratio of a horizontal Butterfly antenna with a noiseless LNA parameterized by its height above a lossy ground with $\epsilon_\text{ground}=$~13 and $\sigma_\text{ground}=$~\SI{5}{\milli\siemens\per\metre}.}
\label{NECsimu}
\end{center}
\end{figure}

As the longest wavelength is around 200~m, the lossy ground is in the near field of the antenna, implying losses that can not be neglected. The finite ground conductivity $\sigma_\text{ground}$ and ground relative permittivity $\epsilon_\text{ground}$ imply ground ohmic losses, represented by the loss resistance $R_\text{loss}$ of Eq.~\ref{sig2noiseequ} at ambient temperature $T_{0}$. In this paper, simulations of antenna gain and antenna impedance are performed with the NEC4 engine using the moment method with far field conditions. $\sigma_\text{ground}$ is set to \SI{5}{\milli\siemens\per\metre} and $\epsilon_\text{ground}$ is set to~13, which are typical values of an ``average'' ground. From Eq.~\ref{sig2noiseequ}, even for a noiseless LNA with $v_\text{lna}=0$, the intrinsic SNR of the antenna is not infinite and depends on the ground losses. To guarantee a SNR of at least 10~dB, $T_{0}\,R_\text{loss}$ should be kept ten times lower than $T_\text{atm}\,R_\text{rad}$. In order to lower ground losses, one could imagine to place the antenna at a 2~m height above a metallic mesh laid on the ground, but to be efficient this mesh would have to be huge, with a surface in the range of one wavelength squared (\SI{\sim9\times10^4}{\metre\squared}). Another solution consists in moving away the antenna from the lossy ground. Then, as the antenna height increases, $R_\text{loss}$ decreases and the signal to noise ratio is increasing, as illustrated in Fig.~\ref{NECsimu}. For this simulation where the LNA noise is set to zero, our criterium consists in choosing a minimum height of the antenna giving a minimum signal to noise ratio of 10~dB. It is fullfilled for a height of 9~m in the worst case of the antenna in the horizontal polarization and at the lowest (2~MHz) observing frequency. Thanks to the choice of short 1.2~m length dipole, antenna weight is minimized, easing the possibility to place it on a 9~m mast by minimizing the mechanical constraints.

\begin{figure}[!ht]
\begin{center}
\subfloat{
  \includegraphics[width=0.5\textwidth]{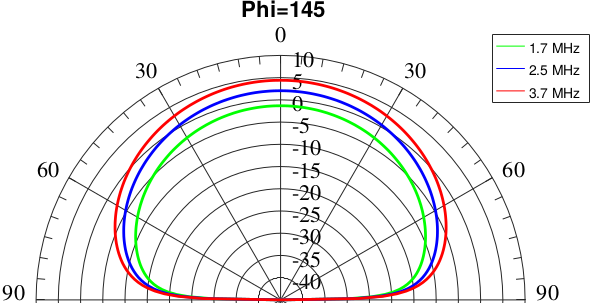}
}
\subfloat{
  \includegraphics[width=0.5\textwidth]{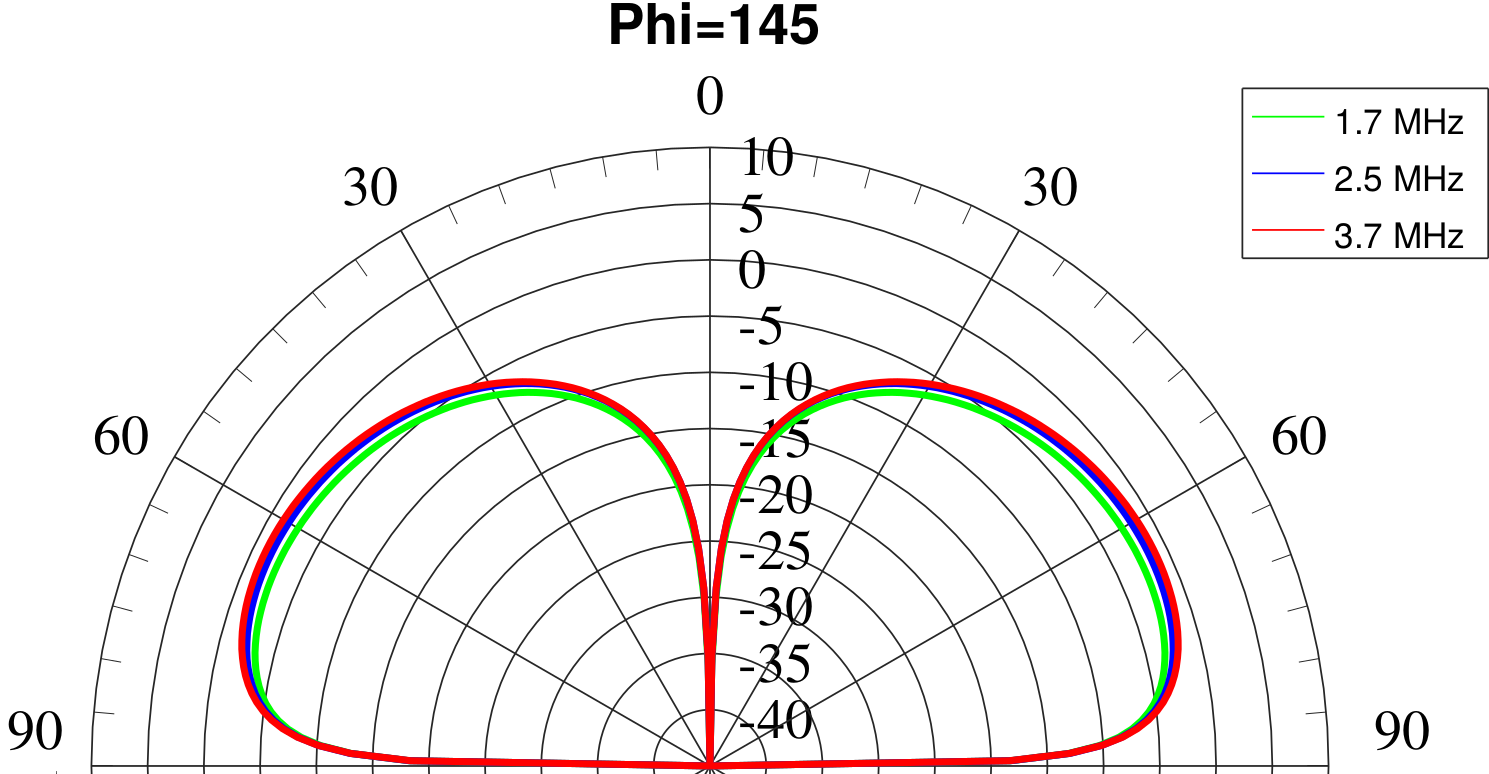}
}
\\
\subfloat{
  \includegraphics[width=0.5\textwidth]{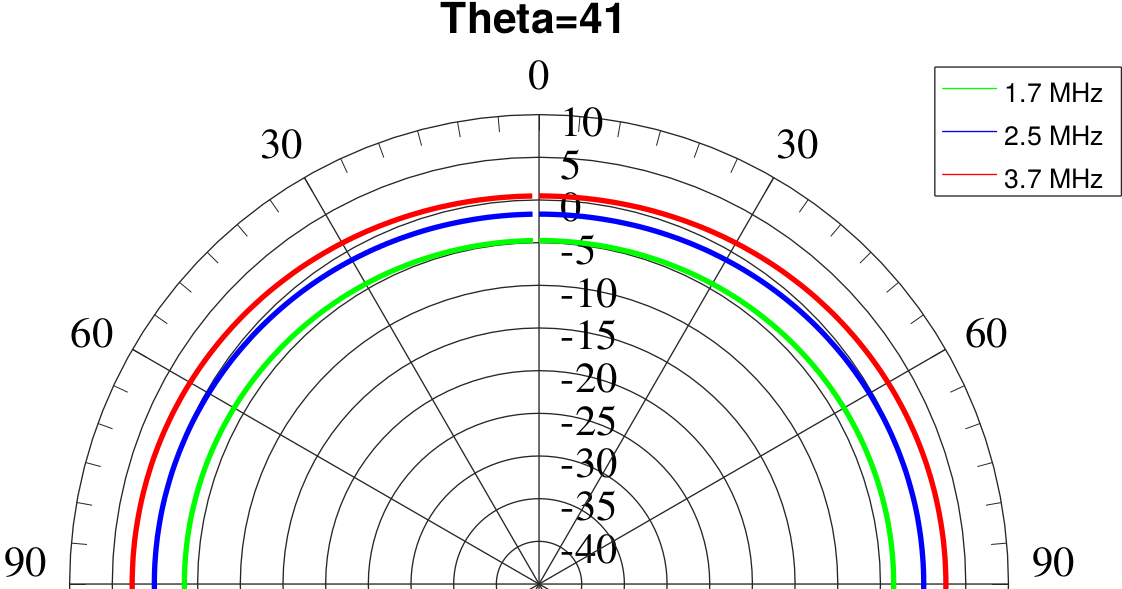}
}
\subfloat{
  \includegraphics[width=0.5\textwidth]{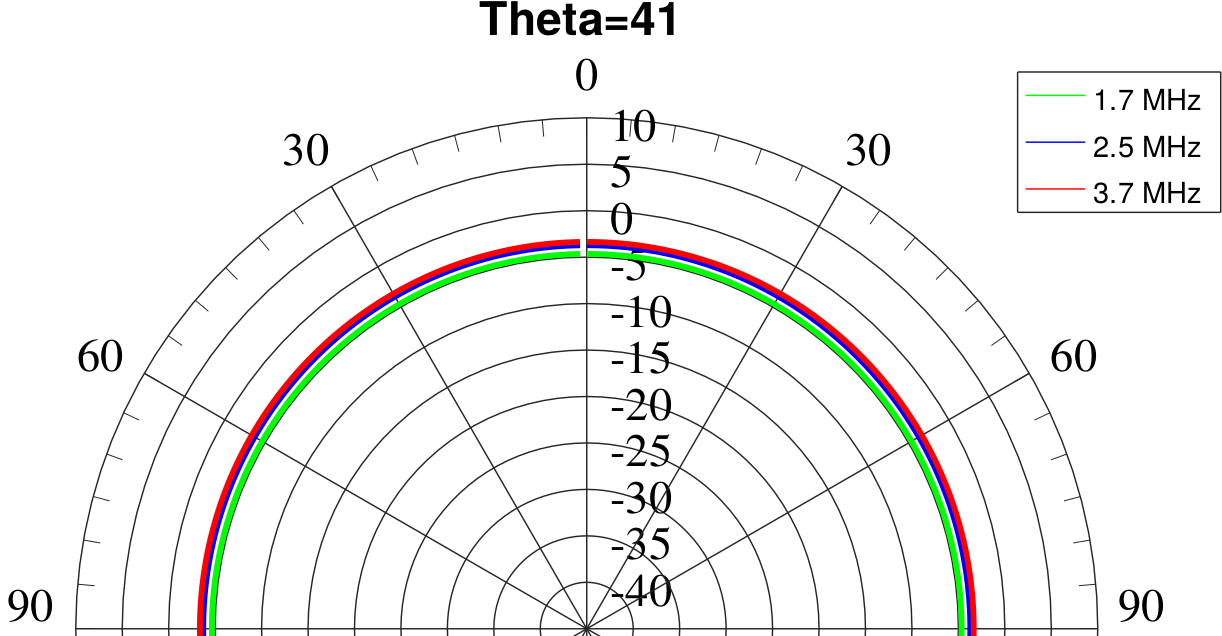}
}
\caption[Simulation of gain pattern of EXTASIS antennas]{Simulation of gain pattern of EXTASIS antennas at 9~m height, as a function of $\Phi$, $\Theta$ and frequency. Top left:  $\Phi=\SI{145}{\degree}$ for the horizontal antenna; bottom left: $\Theta=\SI{41}{\degree}$ for the horizontal antenna; top right: $\Phi=\SI{145}{\degree}$ for the vertical antenna;  bottom right: $\Theta=\SI{41}{\degree}$ for the vertical antenna. Frequencies: green \SI{1.7}{\mega\hertz}, blue \SI{2.5}{\mega\hertz} and red \SI{3.7}{\mega\hertz}. $\Theta=\SI{0}{\degree}$ corresponds to zenith, $\Phi=\SI{0}{\degree}$ to the East and $\Phi=\SI{90}{\degree}$ to the North.}
\label{gainLF}
\end{center}
\end{figure}

Fig.~\ref{gainLF} presents the $9$~m high antenna total gain patterns as a function of the zenith angle and azimuth angles for different frequencies and for both the horizontal and vertical antenna. For these polarizations, and at a fixed azimuth angle, the gain pattern is maximal for a zenith angle corresponding to a vertical direction of arrival and decreases as the zenith angle increases. The horizontal antenna is optimal to detect the LF counterpart of the radio signal emitted during the shower development. Inversely, the vertical antenna is optimal to detect the sudden death signal coming from the ground level, thus the gain pattern is maximal for large zenith angles. At a zenith angle of \SI{41}{\degree}, the gain pattern is homogeneous over the whole azimuth angle range, with a difference of about $3$~dB between the gain of both antennas at $2.5$~MHz. Fig.~\ref{NECsimu} shows that lower heights give lower SNR values: therefore, if the antenna were placed at lower heights, the antenna gain pattern should be corrected by the same offset values. For instance, for an horizontal antenna, the zenith gain of $2.5$~dB at $9$~m would become $-6.5$~dB at $4$~m, and $-19$~dB at $1.5$~m. 

\subsection{From the antenna to the ADC}

At frequencies below $1.7$~MHz the power spectrum is highly dominated by a strong local AM $162$~kHz transmitter and medium waves [526.5-1606.5]~kHz AM transmitters. To allow the LNA to work in its linear zone, far enough from its compression point in daytime conditions, front end high pass and notch filters (not shown in Fig.~\ref{sketch}) are added at the LNA inputs. More exactly than previously mentioned, the input impedance of the LNA is equivalent to a capacitance ($C_\text{lna}$) in parallel to a resistance, thus defining a first order high pass filtering. The settings of the LONAMOS are performed so as the resistance is $10$~k$\Omega$, giving a $900$~kHz cut-off frequency. A passive second order LC high pass filter is also placed in front of the LNA in addition to a LC $162$~kHz notch filter. These three filters give a total attenuation of $53$~dB at $162$~kHz. As shown in Fig.~\ref{sketch}, the output signal of the LNA is transmitted by a $12$~m coaxial cable to a RF analog chain followed by an $8$ bits commercial digital oscilloscope controlled by a dedicated acquisition software hosted by a local PC. The analog chain is composed of a bias-T, allowing to power the LNA via the signal cable, followed by a band pass filter with $600$~kHz and $5.8$~MHz cut-off frequencies in order to attenuate the strong local $162$~kHz transmitter and medium wave transmitters. This chain is ended by an impedance transformer with a 1:8 impedance ratio stuck to the ADC input connector. It performs a 50~$\Omega$ matching (input reflexion coefficient lower than $-18$~dB) to the high impedance of the ADC input in a [$230$~kHz~-~$13$~MHz] bandwidth, and adds a $9.7$~dB voltage gain. Despite the RF signal is bandwidth limited to less than $6$~MHz, the signal is oversampled \cite{ADC} to \SI{500}{\mega \siemens\per\second}
 in order to obtain $14$~dB additional dynamic on the ADC thanks to a digital Finite Impulse Response (FIR) filtering applied in the acquisition software. Consequently, the $8$-bit, $2$~ns sampling digitizer is equivalent to a $10.3$~bits digitizer but with a $10$~MHz limited bandwidth. The FIR filter is a $201$ coefficients gate weighted by a cardinal sine function, in order to obtain an extremely flat magnitude response up to $6$~MHz and a minimum rejection of $-40$~dB from $10$~MHz to beyond. During the FIR filtering, the DAQ software also performs a decimation by a factor of 25 resulting in a final record with a time resolution of 50~ns. Thus, lighter files are stored without any loss of information in view of the Nyquist condition. Oversampling factor, decimation factor and FIR filter coefficients can be changed in the DAQ software.    

\subsection{Trigger signal and acquisition}
\label{triggersignal}
When at least any 5 out of the 13 scintillators of the CODALEMA particle detector are triggered within a given time window, called a ``particle coincidence window'', a master trigger is built and sent to a nearby dedicated GPS station which dates the event at the ns precision, and to the EXTASIS LF antennas. The cosmic-ray energy threshold of our trigger setting has been evaluated between $5\times10^{15}$ and \SI{10^{16}}{\electronvolt}. When requiring offline that at least one MF signal is recorded in coincidence with the scintillators, this energy threshold increases to \SI{5\times10^{16}}{\electronvolt} \cite{2009APh....31..192A}. Due to the extent of the array (several hundreds of meters), the trigger for the LF antennas is distributed over an Ethernet network, which takes on average \SI{\num{750\pm250}}{\micro\second} to reach the antenna. The trigger signal received at each LF antenna is also dated allowing to correct for the network time jitter. The raw digitizer sampling time bin is 2~ns, the recorded trace contains $10^6$ time bins, corresponding to \SI{2}{\milli\second} signal length, long enough to find the particle detector trigger time in the trace by subtracting the trigger emission time from the trigger reception time. Taking into account the extent of the LF antenna array around the particle detector and the propagation speed of the signal (approximated to the speed of light), if any LF radio transient has been recorded in coincidence with the air shower that has triggered the particle detector, it should be located within a \SI{3}{\micro\second}-wide time window around the reconstructed particle trigger time.

\section{Offline data selection}
\label{Sel}
In the frequency range below \SI{10}{\mega\hertz}, the ambient noise is important, and the shower transients are expected to be often buried in the noise. Signal cleaning requires a high-performance signal processing method. Three detection methods have been developed and tested: simple threshold (minimum bias method), linear prediction coding (LPC) \cite{Bohlin:1973:CTM:1664656.1664657} and a combination of wavelet analysis~\cite{Daubechies} and neural networks. They have been compared in order to select the most efficient one. Our set of events contains two days of data recorded during winter, corresponding to \numprint{2535} events triggered by the particle detector. To compare the three methods cited above, a fake transient with a known shape and a known position but with a random amplitude has been systematically added to our raw data. The signals are filtered in the range \SI{[1.7-3.7]}{\mega\hertz} and then, the three methods have been applied to each event. This test is only intended to select the most efficient method, that is why we do the test on a set of events recorded under the worst conditions of background noise. 

The LPC method appears to be the most efficient for finding a LF pulse in our data. In this method, the sample $n$ of a recorded time serie can be modelized as a linear combination of the $n-1$ previous samples:
\begin{equation}
\label{LPCequation}
 s_p(n)=\sum\limits_{k=1}^n a_ks(n-k) 
\end{equation}
where $a_k$ are the predictor coefficients. We call prediction error the difference between the signal $s(n)$ and the predicted signal $s_p(n)$. In the present case of a search for a transient signal in a noise composed of the contribution of several transmitters and the atmospheric noise, the prediction error represents well the expected transient signal. The predictor coefficients are determined by minimizing the sum of squared differences between the true samples and the linearly predicted samples. For more information on the determination of the predictor coefficients, please refer to \cite{1451722}.

\begin{figure}[!ht]
\begin{center}
\subfloat[Filtered \SI{[1.7-3.7]}{\mega\hertz} LF signals.]{
  \includegraphics[width=0.485\textwidth]{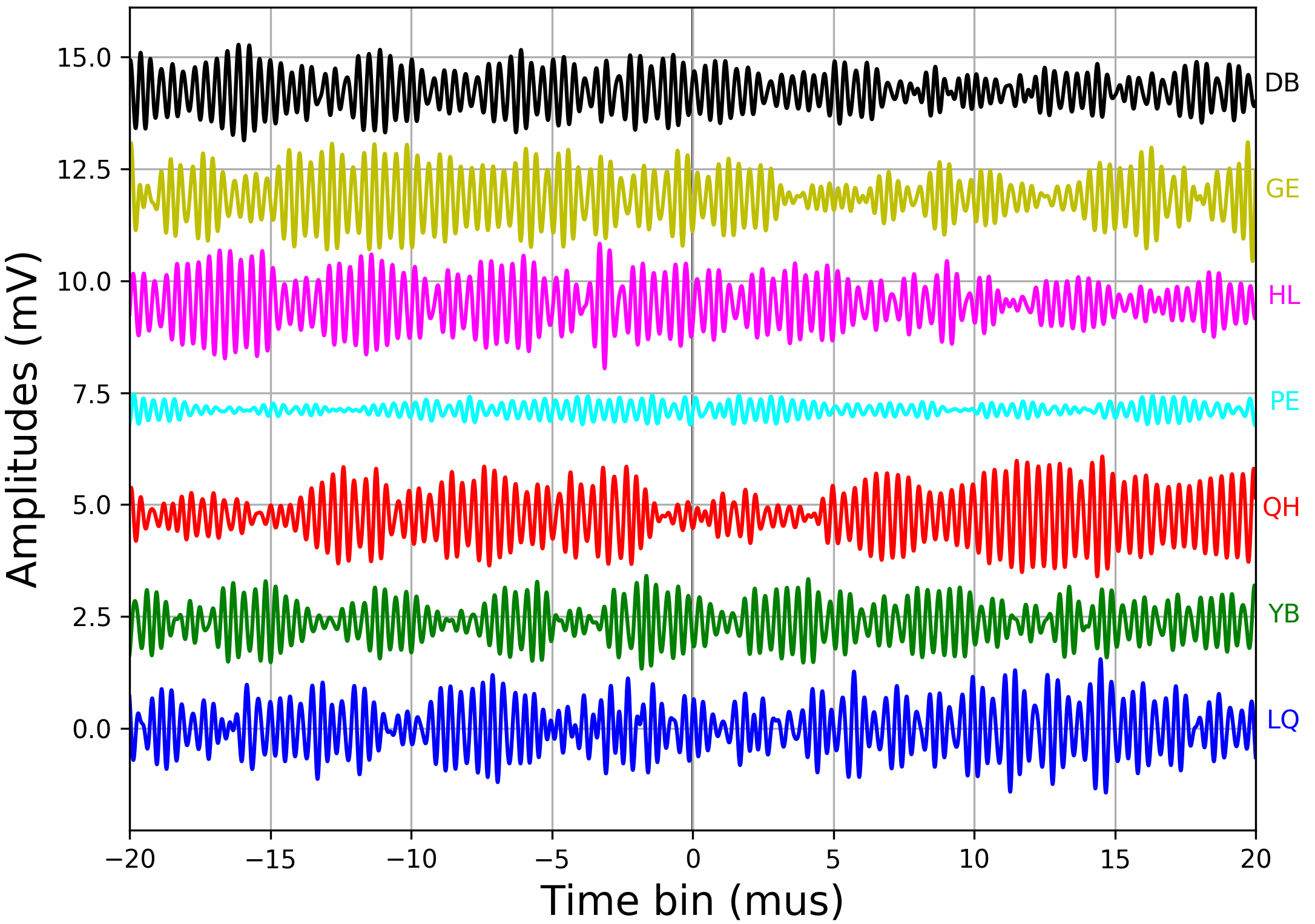}
}
\subfloat[Error of prediction on LF signals.]{
  \includegraphics[width=0.49\textwidth]{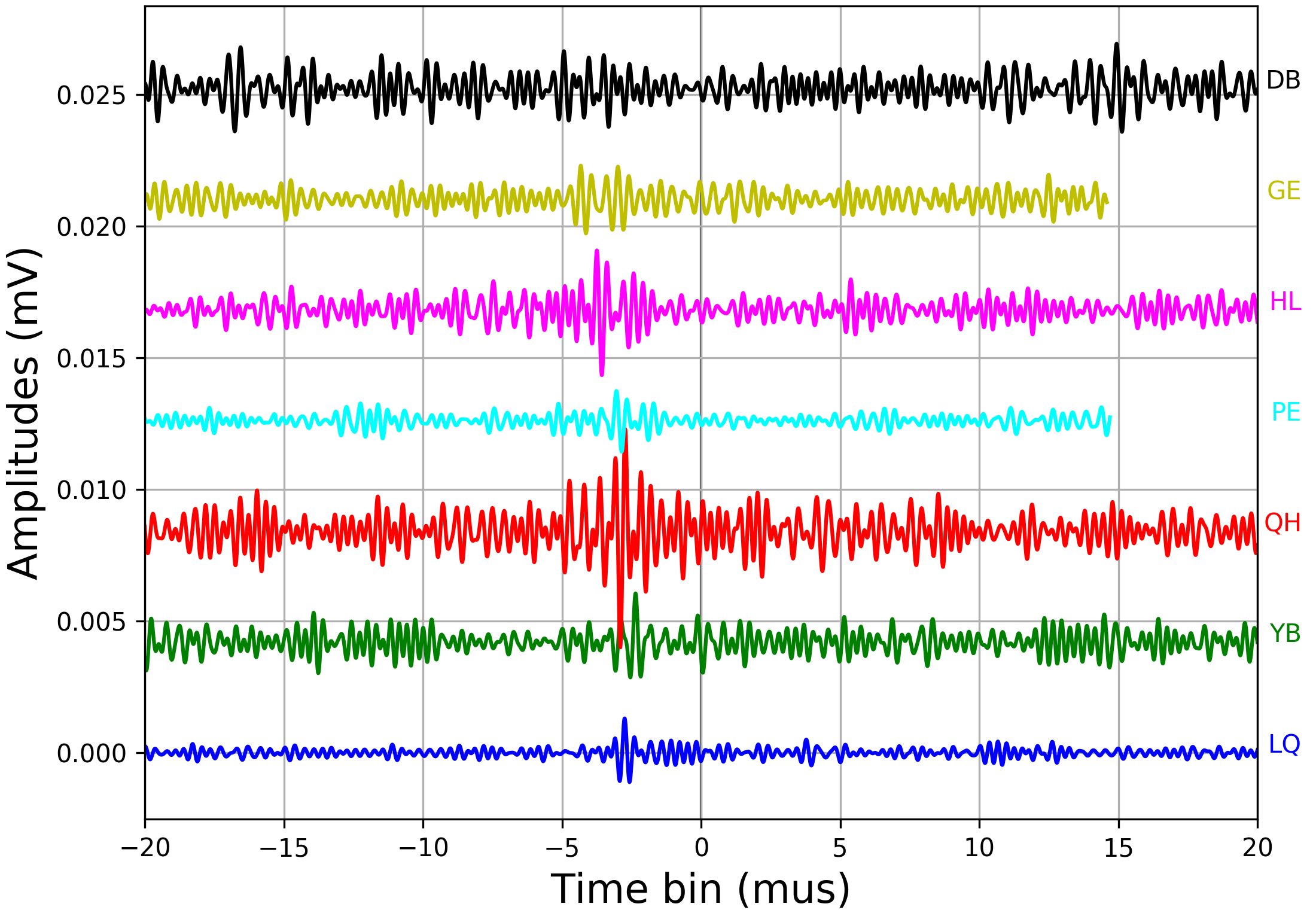}
}
\caption[LF event seen with the EXTASIS instrument]{LF events seen in the horizontal polarizations. Traces are positively shifted along $y$ for clarity. Left: LF signals, as a function of time, filtered in \SI{[1.7-3.7]}{\mega\hertz}, ordered by time of arrival of the cosmic ray signal in antennas. Right: error of prediction of LF signals. Actual cosmic ray transients are detected on traces 3, 4, 5 and 7 from top, in the time window \SI{[\numprint{-5};\numprint{0}]}{\micro\second} after applying a simple threshold method on the LPC prediction error. DB (trace 1) did not detect the transient. Transients visible in traces 2 and 6 are not detected by the LPC method and are located at a time not compatible with the shower geometry.}
\label{secondevt}
\end{center}
\end{figure}

Then, we define a selection threshold as $\max(\mathrm{se})\geqslant\mu_{\mathrm{se}}+K\sigma_{\mathrm{se}}$, where $\mu_{\mathrm{se}}$ and $\sigma_{\mathrm{se}}$ are respectively the mean of the squared prediction error (se) and the standard deviation of the squared prediction error. $K$ is a factor empirically defined as 14 \cite{valcares:tel-00495383}. The result of the procedure is illustrated in Fig.~\ref{secondevt}, where actual detected signals are shown. The transients are not visible on the filtered traces (Fig.~\ref{secondevt}-(a)), but they appear after the LPC processing (Fig.~\ref{secondevt}-(b)). 

\begin{figure}[!ht]
\begin{center}
  \includegraphics[width=0.8\textwidth]{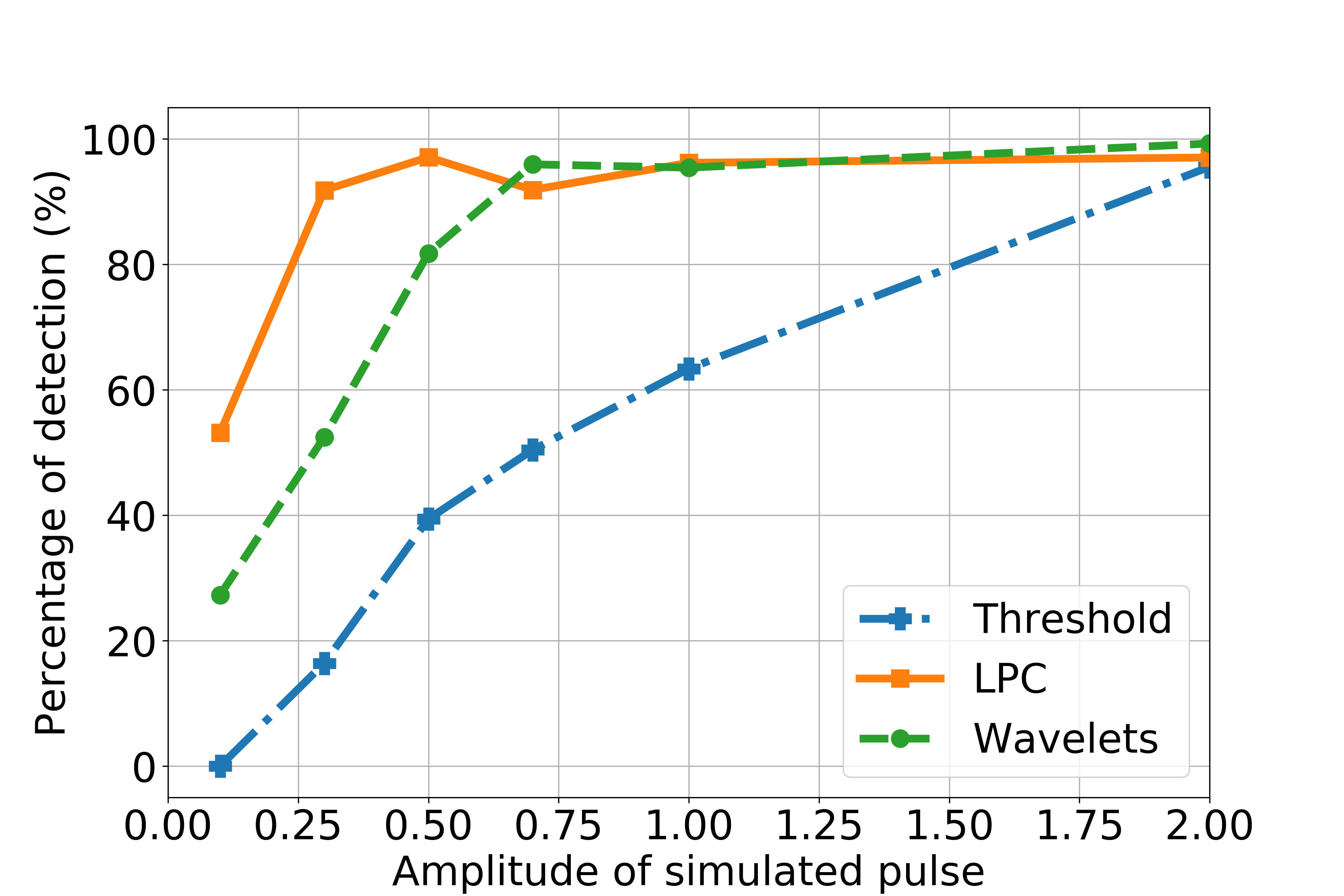}
\caption{Detection efficiency of the three tested methods, as a function of the amplitude of the simulated pulse relative to the observed noise peak to peak value. Plain lines is for the LPC method, dashed line for the threshold method and dotted-dashed line for the wavelet method.}
\label{DetPerH}
\end{center}
\end{figure}

The LPC and wavelet methods have better detection efficiencies at low pulse amplitude compared to the threshold method, as shown in Fig.~\ref{DetPerH}. The gap decreases for high amplitude pulses, but the LPC and wavelet methods are still the most efficient. By also comparing the percentage of false detection, we found around \SI{\numprint{4}}{\%} for the LPC method and \SI{\numprint{10}}{\%} for the wavelet method. These are the reasons why the LPC method was chosen. 

\section{Results}
\label{fresults}
In order to reach the EXTASIS objective to detect and exploit LF signals, data from the CODALEMA instruments and EXTASIS were combined. In the following we first present how a complete cosmic-ray event is detected and analysed, and then we illustrate a LF signal detection. 

\subsection{Event reconstruction}
\label{Coincidences}

This section aims to describe the selection procedure of the events seen simultaneously by multiple instruments of CODALEMA and EXTASIS, and the elaboration of a so-called ``cosmic ray event'' associating several of these instruments. As it has been discussed in section \ref{triggersignal}, the trigger is generated by the particle detector and distributed to the compact array and to the LF antennas. For the 57 standalone antennas, no particle trigger is sent. Transients detected in coincidence on several of these standalone antennas build a ``radio coincidence'', characterized by an average radio event time that will be compared to the particle detector event. The criterion is that at least three standalone antennas are triggered within a time interval compatible with the propagation of a plane wave at the speed of light. The radio event is promoted as an actual shower if its timing is compatible with the timing of the scintillators and if the reconstructed arrival directions agree within \SI{20}{\degree}~\cite{CODA3}.

Since the installation of the complete instrumental setup of EXTASIS (March 2017) and until the end of year 2017, 767 cosmic events have been detected by the CODALEMA standalone antennas, with a potentially detectable LF signal. Among them, 446 present a pulse detected by the LPC method on at least one LF antenna. We checked that LF events with a multiplicity of 1 and 2 are mostly fortuitous, their detection time being incompatible with the expected time deduced from the MF signal reconstruction. Moreover, LF events with a multiplicity lower than 3 are not interesting for us in this study, where we aim to know whether the LF counterpart can improve the shower reconstruction or not. We thus decided to ignore LF events with a multiplicity lower than 3, knowing also that a large majority of them are probably accidentals, taking into account the transient noise rate and the wide time window of search. For LF events with a multiplicity of 3, the coincidence rate falls dramatically, since only 18 of these 767 events present a LF counterpart according to the LPC method (see section~\ref{discussion}). We do not apply any angular criterion on the selection of the LF events, because the uncertainties on the directions of arrival are estimated around \SI{\sim10}{\degree} (compared to \SI{\sim1}{\degree} for the standalone antennas), mostly due to the transient duration in the LF band.

These events are summarized in table~\ref{lfeventtable}. They have a timing compatible with the usual radio emission when the shower develops in the air and not with what we expect from the shower sudden death. Unfortunately, the next reconstruction steps (core location, $X_{\mathrm{max}}$ and energy estimation from the MF signals) can not be performed for most of these events for various reasons:
\begin{itemize}
 \item a core location clearly well outside of the CODALEMA array requires an hazardous extrapolation of the electric field predicted by SELFAS3,
 \item a low multiplicity leads to a very poor accuracy on the shower parameters,
 \item an abnormal and undetermined atmospheric electric field kills the correlation between the electric field measured in the antennas and the shower development.
\end{itemize}
As a consequence, the fact that the energy can not be estimated in the EMF band for these eighteen events does not allow to deduce properties on the amplitude of the LF signal. The unique event (number 4) for which the reconstruction appears reliable is detailed in the following section.

\begin{table}[!ht]
\centering
\small
\begin{tabular}{|c|c|c|c|c|c|c|}
\hline
$\#$ & $N_\mathrm{SA}$ & $\theta_{\mathrm{SA}}$ (\si{\degree}) & $\phi_{\mathrm{SA}}$ (\si{\degree}) & $\Delta\alpha$ (\si{\degree}) & $E_{\mathrm{stat}} (\si{\kilo\volt\per\metre})$ & Probability \\ \hline
1  & 6  & 31.8 & 353.1 & 2.1   & 7.5   & $1.91\times10^{-3}$\\ \hline
2  & 20  & 60.0 & 153.8 & 6.0   & 0.3  & $2.85\times10^{-1}$ \\ \hline
3  & 11  & 28.8 & 68.9 & 3.2   & 1.8  &  $6.35\times10^{-3}$ \\ \hline
4  & 11  & 40.6 & 145.2 & 11.3  & -0.1 &  $1.14\times10^{-1}$ \\ \hline
5  & 13  & 40.9 & 32.6 & 20.4  & -0.1 & $1.14\times10^{-1}$ \\ \hline
6  & 8  & 56.1 & 58.9 & 3.4   & 0   & $7.42\times10^{-1}$ \\ \hline
7  & 3  & 34.2 & 252.9 & 3.8   & 0.3  & $2.85\times10^{-1}$ \\ \hline
8  & 4  & 53.4 & 95.9 & 13.3  & 0.1  & $4.97\times10^{-1}$ \\ \hline
9  & 4  & 44.4 & 76.6 & 22.9  & 0.1  & $4.97\times10^{-1}$ \\ \hline
10 & 7  & 16.2 & 210.8 & 6.3   & -12.3 & $1.19\times10^{-3}$ \\ \hline
11 & 25  & 38.4 & 206.5 & 3.4   & -15.7 & $5.09\times10^{-4}$   \\ \hline
12 & 3  & 77.7 & 14.5 & 9.1   & 0.1  & $4.97\times10^{-1}$ \\ \hline
13 & 5  & 24.5 & 235.3 & 3.7   & -0.5 & $3.23\times10^{-2}$ \\ \hline
14 & 22  & 23.0 & 92.3 & 6.0   & -11.1 & $1.51\times10^{-3}$ \\ \hline
15 & 21  & 46.5 & 109.8 & 5.0   & -3.7 & $7.31\times10^{-3}$ \\ \hline
16 & 20  & 25.9 & 74.3 & 14.1  & -7.6 & $3.07\times10^{-3}$ \\ \hline
17 & 24  & 33.5 & 23.4 & 10.3  & 2   & $5.25\times10^{-3}$ \\ \hline
18 & 8  & 7.4 & 203.8 & 1.5   & -6.5 & $2.28\times10^{-3}$ \\ \hline
\end{tabular}
\caption{List of the 18 LF events selected. The first column is the event identification number. The second one gives the multiplicity of the standalone antennas of the event. The third and fourth columns give respectively the zenith and azimuth angles of the event reconstructed by the standalone antennas. The fifth column gives the 3D-angular difference of the arrival direction of the event reconstructed by the MF and LF antennas. The sixth column indicates the value of the atmospheric electric field recorded during the detection of the event, and the corresponding probability (see section~\ref{atmelec}).}
\label{lfeventtable}
\end{table}

\subsection{Example of low-frequency signal detections}
\label{LFdet}
\begin{figure}[!ht]
\begin{center}
	\begin{tikzpicture}
	\node[anchor=south west,inner sep=0] (image) at (0,0) {\includegraphics[width=0.65\textwidth]{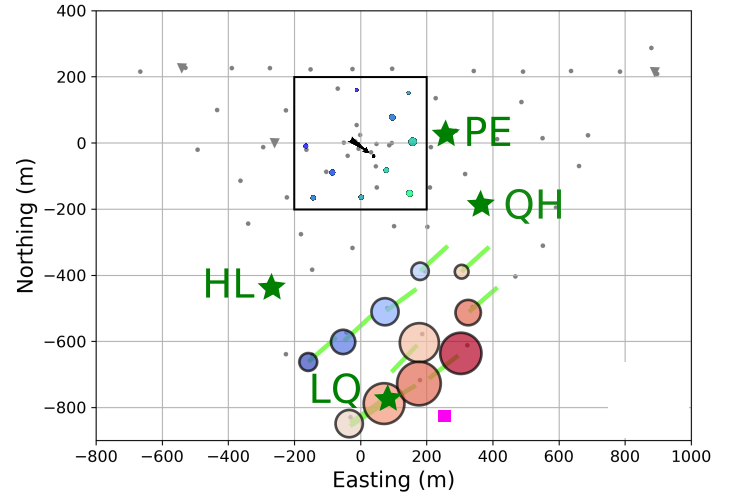}};
	\begin{scope}[x={(image.south east)},y={(image.north west)}]
    \draw[red,line width=1mm,<-] (0.31,0.48) -- ++(canvas polar cs:angle=145,radius=1cm);
	\end{scope}
      \end{tikzpicture}
\caption[Map of a part of facilities in Nan\c cay with the involved instruments]{Footprint of the event~4 seen by the particle detector and some LF and MF antennas. The arrival direction is represented by the red arrow, and its energy \SI{(3.7\pm0.6)\times10^{18}}{\electronvolt}, from the CODALEMA standalone antenna reconstruction. Grey dots represent the standalone antennas, the square area represents the scintillators array region. The green stars indicate the positions of the LF antennas that detected the shower. The involved standalone antennas are represented by coloured circles, whose colour indicates the timing order in which the signal has been seen by the antennas (from blue, earliest, to red, latest) and area of circles reflects the relative amplitude of the signal (linear scale). The small green lines close to the circles indicate the orientation of the measured polarization of each MF antenna, nearly orthogonal to the direction of arrival of the event as expected from the dominant geomagnetic mechanism. The estimated shower core location is represented by the magenta square ($x=\,$\SI{259\pm35}{\metre} and $y=\,$\SI{-809\pm30}{\metre}). See text for more details on the cosmic ray and shower properties of this event.}
\label{mse}
\end{center}
\end{figure}

A ground map of the event 4 of table \ref{lfeventtable} is shown in Fig.~\ref{mse}. Eleven standalone antennas (circles) at the south part of the MF array have recorded a signal. LF counterparts were registered in four LF antennas. The small green lines close to the circles indicate the orientation of the polarization of each MF antenna, expected to be nearly orthogonal to the direction of arrival of the event. The recorded traces are shown in Fig.~\ref{secondevt}, ordered by time of arrival in the LF antennas. The pulses located around \SI{-2.5}{\micro\second} on GE and YB antennas ($2^\text{nd}$ and $6^\text{th}$ traces from top) are fortuitous transients, rejected by both the LPC procedure and by the Direction Of Arrival (DOA) reconstruction.

Extensive, iterative and systematic comparisons of the EMF signals with SELFAS3 simulations were performed to deduce the cosmic ray and shower properties. It includes antenna by antenna spectrum comparisons (on both polarizations) over a substantial set of simulations spanning a large range in core position, shower $X_\text{max}$ and cosmic-ray energy (see \cite{gatef,LilianICRC2017} for more explanations on the method). Moreover, the amplitude and spectral index variations observed in the eleven MF antennas are well reproduced by the simulation selected at the end of this iterative process. It gives us a strong confidence on the event reconstruction reliability. The best core position, represented by a magenta square in Fig.~\ref{mse}, is $x=\,$\SI{259\pm35}{\metre} and $y=\,$\SI{-809\pm30}{\metre} (our reference position is located at the center of the particle detector array). The method also gives an estimate of $X_{\mathrm{max}}$ of \SI{715\pm19}{\gram\per\centi\metre\squared} and an energy of \SI{(3.7\pm0.6)\times10^{18}}{\electronvolt}. The latter is in good agreement with the energy estimated by the particle detector of \SI{(2.75\pm1.05)\times10^{18}}{\electronvolt}. This event core location being external to the particle detector area, the shower core location determined by the radio method has been used for the particle detector energy reconstruction. This explains the large uncertainty on the energy estimated with the scintillator data. 

\begin{figure}[!ht]
\begin{center}
\subfloat[\SI{[30-80]}{\mega\hertz}.]{
  \includegraphics[width=0.485\textwidth]{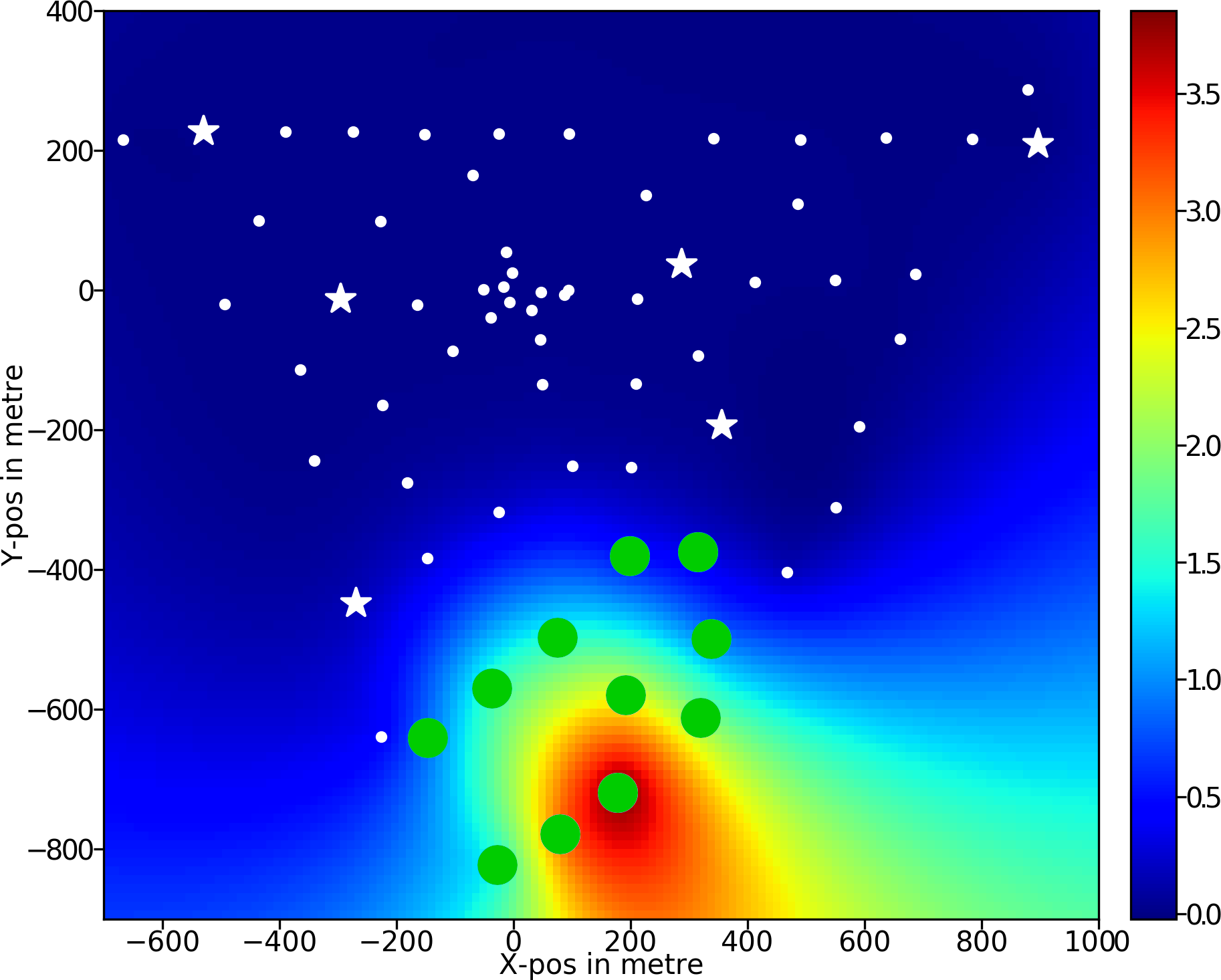}
}
\subfloat[\SI{[1.7-3.7]}{\mega\hertz}.]{
  \includegraphics[width=0.5\textwidth]{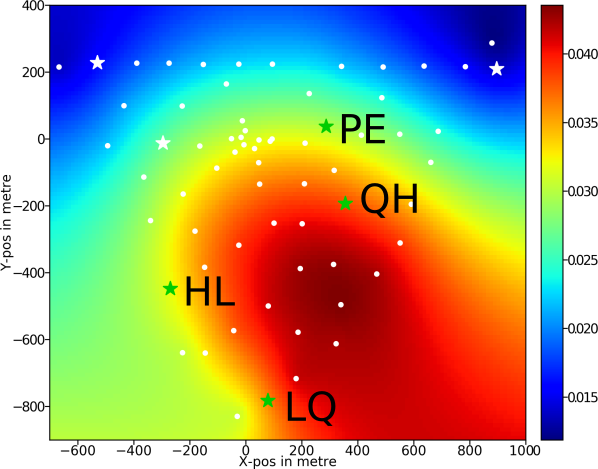}
}
\caption{Lateral distribution of the electric field depending on the frequency range predicted by SELFAS3. Left: \SI{[30-80]}{\mega\hertz}. Green circles represent the involved MF antennas in the event. Right: \SI{[1.7-3.7]}{\mega\hertz}. Stars represent the LF antennas, green ones correspond to the involved LF antennas in the event. The LF antenna (PE) located around ($x=\,$\SI{300}{\metre}~; $y=\,$\SI{20}{\metre}), \SI{850}{\metre} from the shower core location at ground, gives the extent of the detection zone at low frequency. The color scale, expressed in\SI{}{\milli\volt\per\metre}, is not the same for the two plots: the detected electric field in the LF band is smaller.}
\label{ldf}
\end{center}
\end{figure}

Fig.~\ref{ldf} presents the interpolated simulated electric field of the horizontal polarization in \SI{[30-80]}{\mega\hertz} and \SI{[1.7-3.7]}{\mega\hertz}. At low frequency (Fig.~\ref{ldf}, right), the electric field distribution appears much wider and flatter than at medium frequency (left), with a considerably increased detection range. Indeed, the LF antenna PE located around ($x=\,$\SI{300}{\metre}~; $y=\,$\SI{20}{\metre}), see Fig.~\ref{nancay2}, has detected the shower at \SI{850}{\metre} from the reconstructed shower core location, while the most distant MF antenna is only at \SI{400}{\metre} from the latter. There is no MF counterpart in the standalone antenna associated with PE. This hints an electric field detection threshold of about \SI{23\pm4}{\micro\volt\per\metre} at low frequency in the horizontal polarization, the value detected on the PE antenna after correction for the antenna equivalent length and acquisition chain gains: the GE antenna, located at ($x=\,$\SI{-250}{\metre}~; $y=\,$\SI{0}{\metre}) more or less at the same distance of the shower core, has not detected the simulated electric field of \SI{23}{\micro\volt\per\metre}.

Moreover, as it can be seen in Fig.~\ref{ldf} where the color scale is expressed in\SI{}{\milli\volt\per\metre}, the electric field in the LF band is actually smaller than in the EMF band. This result disagrees with the pioneer observations, which reported that when frequency decreases, a clear evidence of a strong increase of the radio pulse amplitude was seen.\\

\begin{figure}[!ht]
\begin{center}
\subfloat{
	\begin{tikzpicture}
	\node[anchor=south west,inner sep=0] (image) at (0,0) {\includegraphics[width=0.5\textwidth]{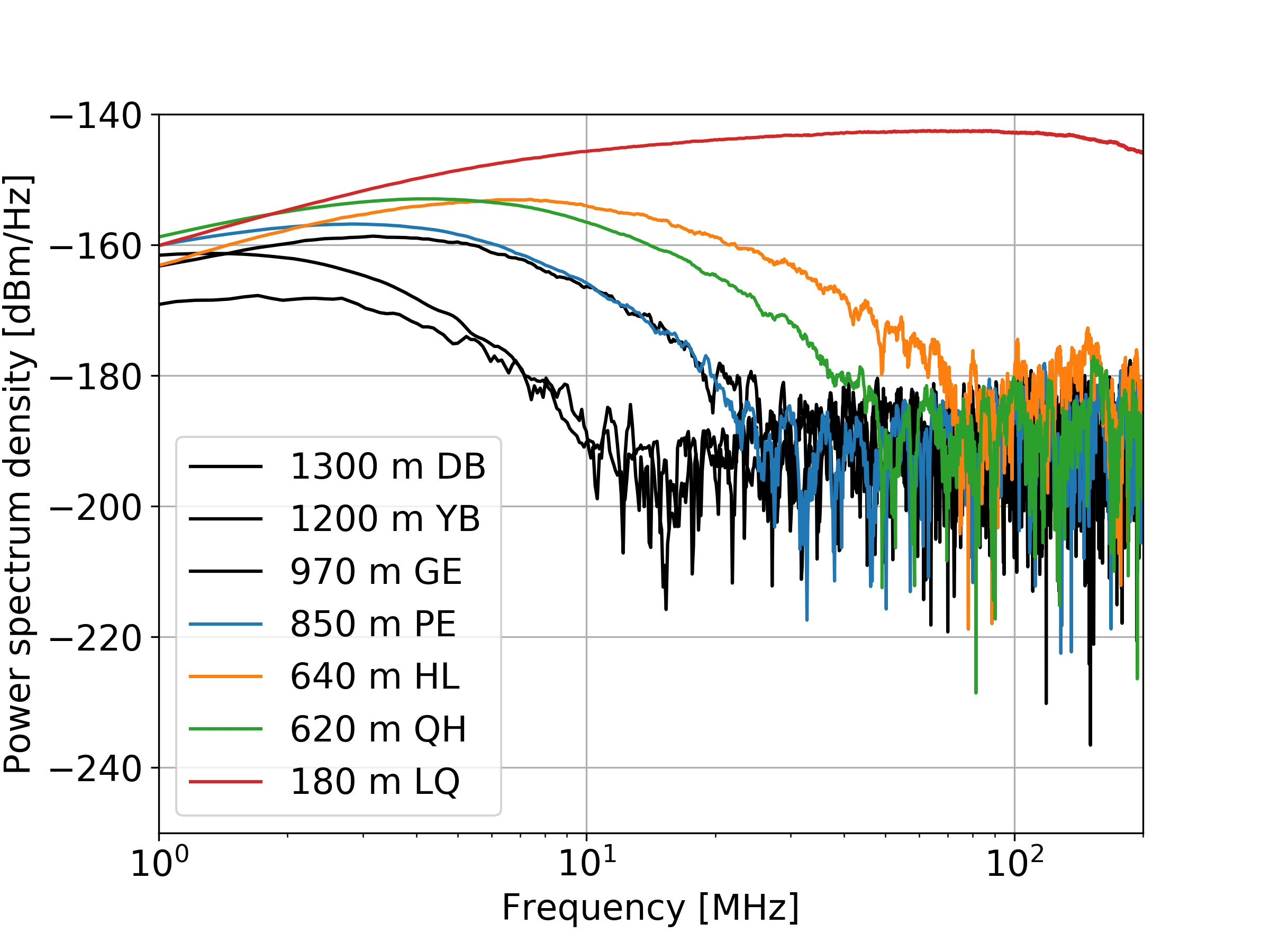}};
	\begin{scope}[x={(image.south east)},y={(image.north west)}]
	\fill[Peach,fill opacity=0.25]
	(0.55,0.88)
	-- (0.9,0.88)
	-- (0.9,0.12)
	-- (0.55,0.12)
	-- cycle;
	
	\fill[LimeGreen,fill opacity=0.25]
	(0.2,0.88)
	-- (0.32,0.88)
	-- (0.32,0.12)
	-- (0.2,0.12)
	-- cycle;	
	\end{scope}
      \end{tikzpicture}
}
\subfloat{
  \includegraphics[width=0.5\textwidth]{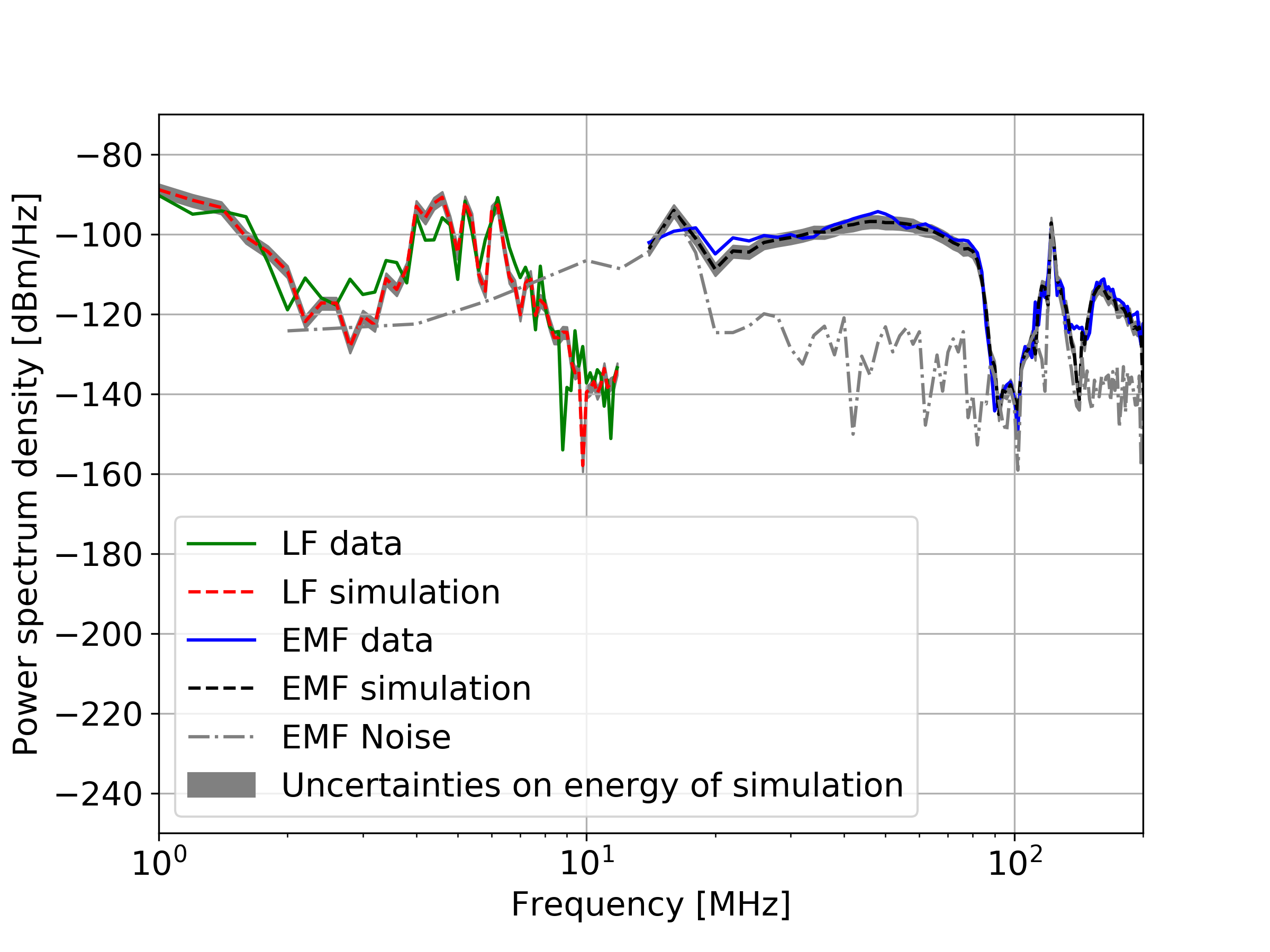}
}
\caption[Simulated power spectrum density as a function of frequency]{Left: simulated power spectrum density as a function of frequency calculated at the LF antenna locations, in colour for the involved LF antennas and in black for the others. Distance to shower axis is also indicated. The green band indicates the range of the LF band, and the peach band indicates the range of EMF band. Right: convoluted power spectrum density as a function of frequency for the southernmost LF and MF antennas, labelled LQ on the left figure.}
\label{psd}
\end{center}
\end{figure}

\begin{figure}[!ht]
\begin{center}
  \includegraphics[width=0.9\textwidth]{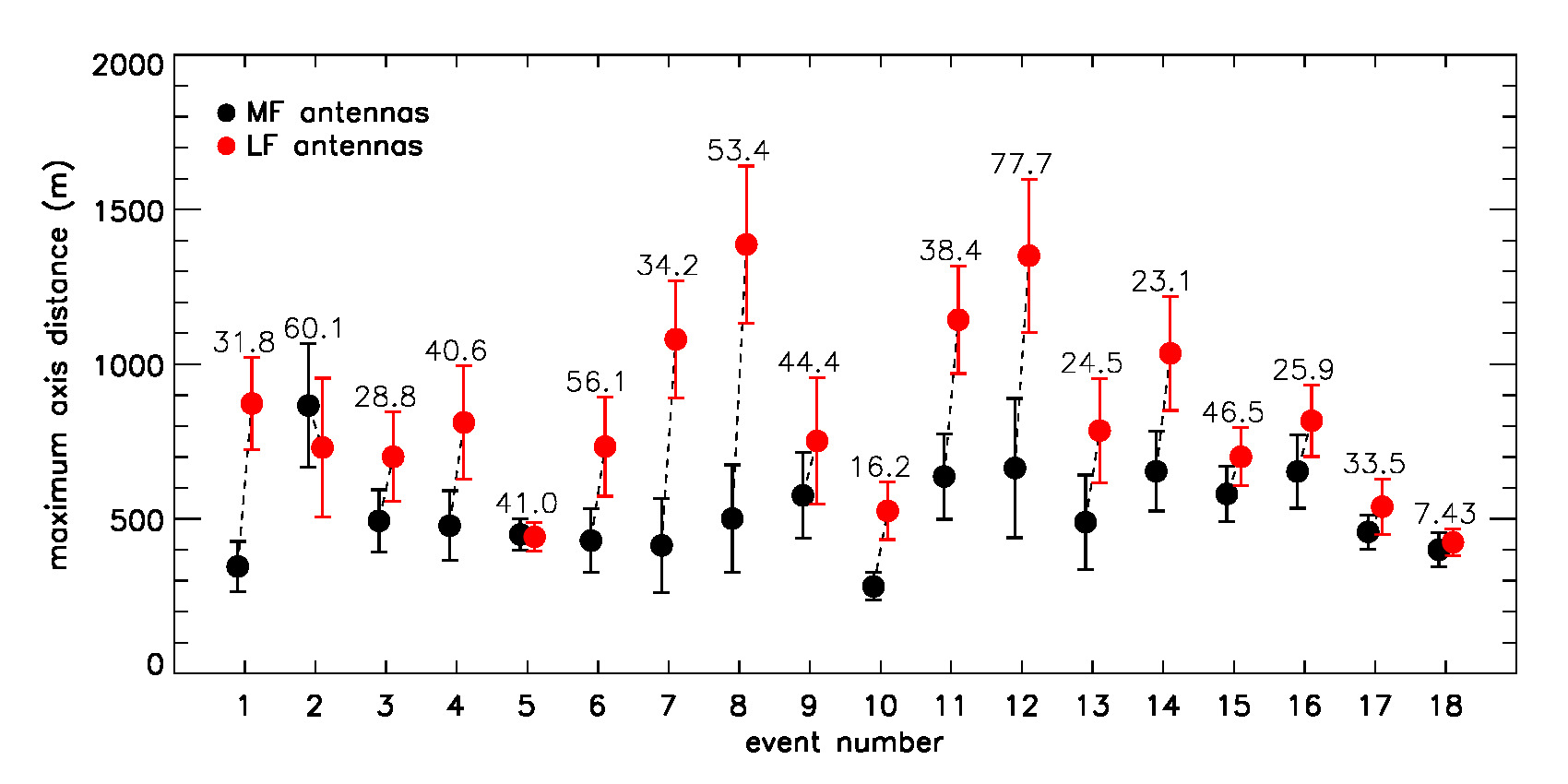}
\caption[]{Maximum axis distances for the LF and MF detectors of the 18 events of table~\ref{lfeventtable}, illustrating that the detection range is higher for LF detectors that the MF detectors. The zenith angles of the events are indicated at the top of the error bars. Large error bars correspond to external events with a large uncertainty on the core position. See text for details.}
\label{MFLFrange}
\end{center}
\end{figure}

Fig.~\ref{psd}-left shows the simulated PSD as a function of frequency and for different LF antenna locations, in color for the involved LF antennas and in black for the others. The PSD quickly drops in the EMF band with the shower axis distance, while it decreases much more slowly in the LF band.
Fig.~\ref{psd}-right presents the PSD of the signal of the shower development over the whole frequency band. LF data are represented by the green line, EMF data by the blue line, and in red and black dashed lines the convoluted simulated power spectrum density in LF and EMF band respectively, in which we have added the noise of the corresponding band. The noise-added, convoluted simulations are in good agreement with the data, showing a good understanding of our LF and MF instruments, but also a good radio reconstruction of the characteristics of the primary cosmic ray.
This result indicates again that the detection range should be larger at the LF band than in the EMF band.
Our data confirm this expectation. One way to quantify the detection range is to consider the axis distance. For a given event, if we know the core position, we can compute the axis distances between each detector and the shower axis. The maximum value of these axis distances is the maximum axis distance for this event and is an indication of the detection range. We don't have a proper core reconstruction for our 18~events but we can define by eye a confidence zone where it should be, as suggested by the ground pattern of the triggered MF detectors. We used such circular zones for each event, with a confidence radius varying from 200~m (for internal events) up to 600~m (for external events). This is a conservative way to get an estimate of the core position and, consecutively, of the axis distances. Then, we compute the average and rms of the maximum axis distance for each event using a large number of sample core positions, taken inside the pre-defined confidence zone. These values are shown in Fig.~\ref{MFLFrange}, in black and red for the MF detectors and LF detectors, respectively. We see that the maximum axis distance is larger for LF detectors than for MF detectors, for almost all events.
This was expected since a long time through the various simulations reported in the literature, but this is the first time that it is confirmed by an actual detection.


At last, for the LF antenna in the MF zone (LQ), and thus at a given shower axis distance, there are \SI{10}{\deci\bel m\cdot\hertz^{-1}} between the maximum in the EMF band and the maximum in the LF band, showing that the signal is larger in the classical band than in the LF band. For the southernmost LF antenna and its MF companion (LQ), the simulated power spectrum density has been convoluted with the antenna and acquisition chain responses to obtain the equivalent in ADC counts, in order to compare them with the raw data.

\section{Discussions}
\label{discussion}
\subsection{How to explain the low detection rate in the low-frequency band?}
\label{lowratediscussion}

\begin{figure}[!ht]
\begin{center}
  \includegraphics[width=\textwidth]{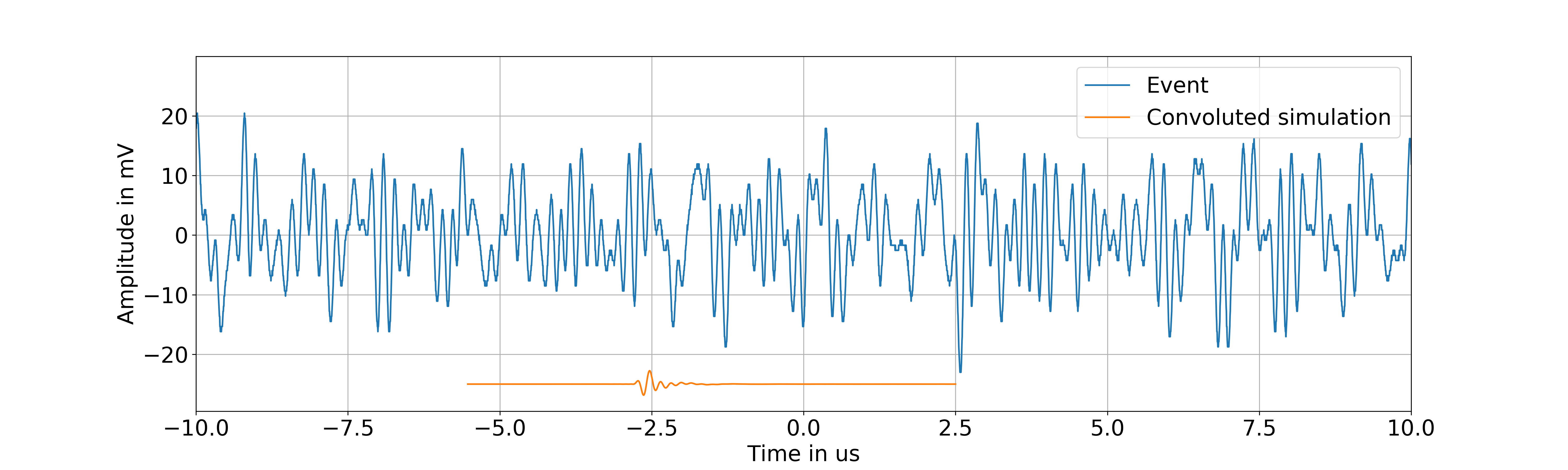}
\caption[]{Blue line: full band signal of the LF antenna closest to the shower core of event of Fig.~\ref{mse}. Orange line: simulation of the signal at the same location and in the same frequency band, convoluted with the antenna response and placed at the time bin where the actual signal has been detected (shifted downwards for visibility).}
\label{LFconvevent}
\end{center}
\end{figure}

As already mentioned in section~\ref{bkgan}, the LF sky is dominated by the atmospheric noise and the noise level at night is $\sim{100}$ times higher than during the day. Consequently, over one year of observation the duty cycle is reduced by a factor of 2. This seasonal variation shown in figures \ref{LFsky}.(e) and (f), which considerably reduces the available daily time, makes a LF detection during winter highly unlikely.\\
Due to the noise, the signal to noise ratio is expected to be much smaller at LF than at MF. As an illustration, let us study the event shown in Fig.~\ref{mse}. For this event, accurate simulations have been carried out as explained in section \ref{LFdet} and the response of the LF antennas has been taken into account, by convolving the simulations. Fig.~\ref{LFconvevent} depicts the detected signals for the southernmost LF antenna (blue curve) and its simulation (orange curve). This antenna has the highest detected signal of the event. The expected signal has been superimposed at the time bin where the actual signal has been detected.

\begin{figure}[!ht]
\begin{center}
  \includegraphics[width=\textwidth]{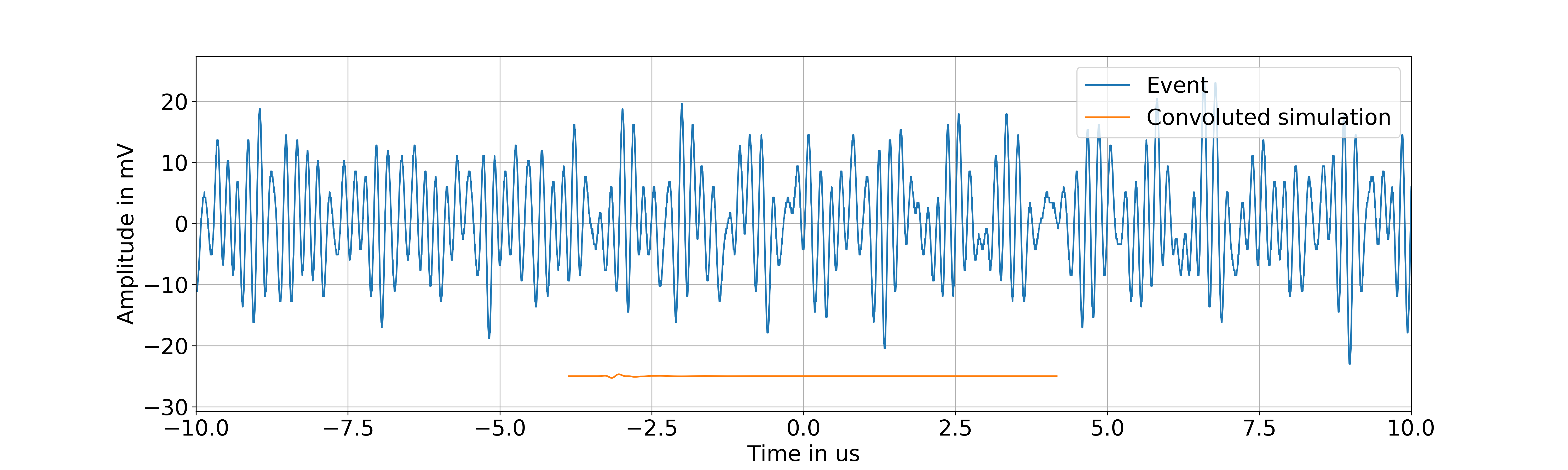}
\caption[]{Same as Fig.~\ref{LFconvevent} for the closest LF antenna without detection. Barely visible, the transient amplitude (around \SI{-3}{\micro\second}) is equal to 1~\% of the background noise RMS.}
\label{LFconvevent2}
\end{center}
\end{figure}

The rms of the background noise is 10 times larger than the amplitude of the convoluted signal, explaining why it is not visible by eye. However, it has been detected using the LPC method. With the LPC method (orange line of Fig.~\ref{DetPerH}), the detection efficiency is around \SI{50}{\%} for an amplitude one order of magnitude below the noise rms. For the considered event, 4 antennas over 7 present a signal after the LPC processing. For the closest of the three antennas without detection, the transient amplitude is estimated from the simulations to be equal to 1~\% of the rms of the background noise (Fig.~\ref{LFconvevent2}). For this amplitude level, the detection efficiency is much smaller than \SI{50}{\%}, explaining why the transient can not be seen even with the LPC method. These observations permit to determine the minimum amplitude of a detectable signal compared to the background noise. In that case, the signal in the farthest LF antenna that has detected the shower has an amplitude of 20~\% of the rms of the background noise.

The low detection rate of LF signals can be thus explained by the atmospheric background noise level which, in the best case, is 10~times higher than the expected minimum detectable signal and makes the detection of the LF counterpart of the shower development unlikely, at least for the shower properties expected at the CODALEMA site. Moreover, as it will be explained in the next sub-section, it is likely that even fewer showers would have been detected if particular atmospheric conditions did not probably amplify the radio signal.

\subsection{Correlation with the atmospheric electric field}
\label{atmelec}

\begin{figure}[!ht]
\begin{center}
  \includegraphics[width=\textwidth]{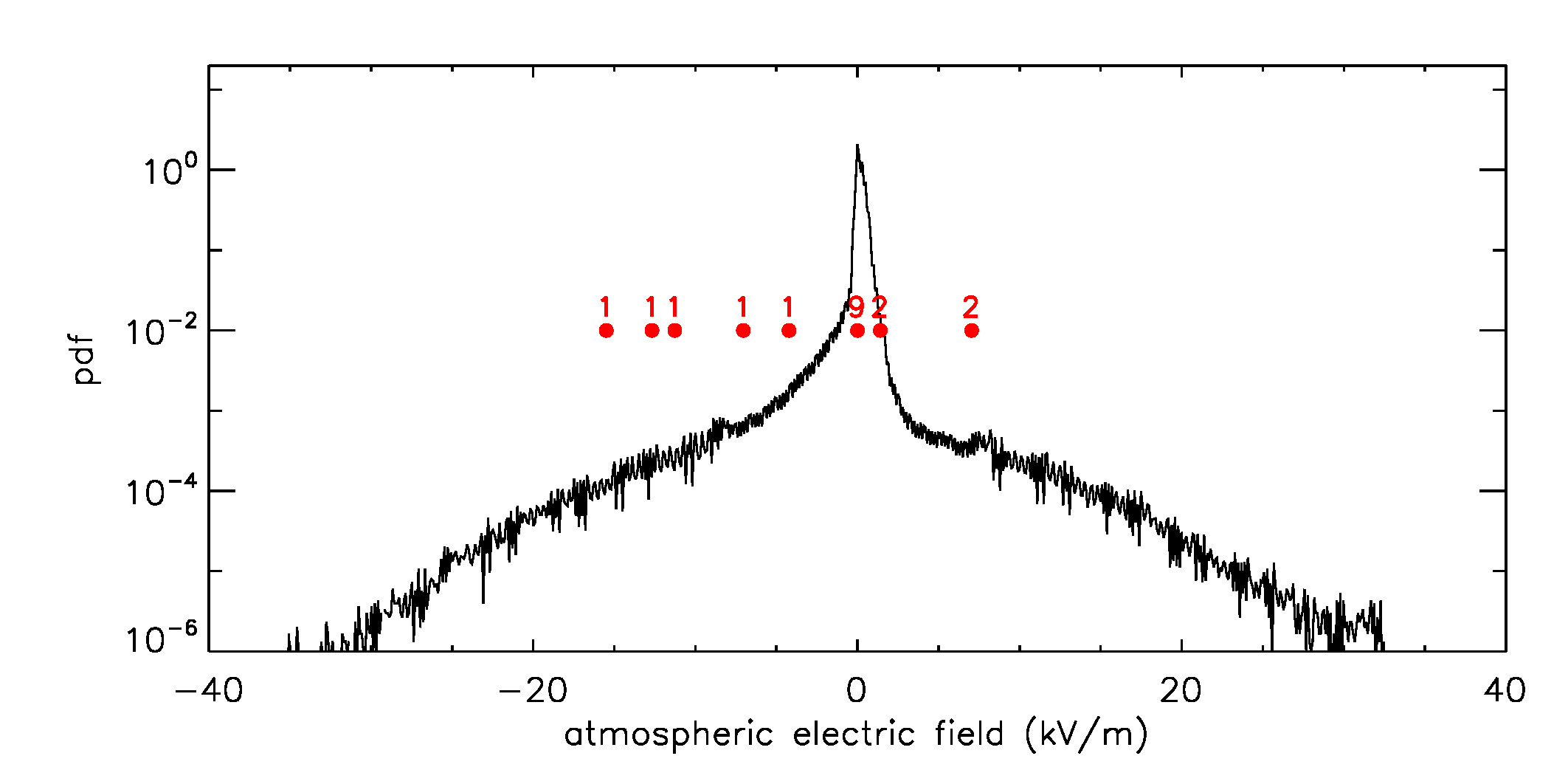}
\caption[]{In black: probability density function of the atmospheric electric field measurements carried out since 2014/09/01, in Nan\c{c}ay (bin size is \SI{100}{\volt\per\meter}). The atmospheric electric field values at the time of the eighteen LF events are represented by the red dots, with the number of corresponding events (bin size is \SI{1.4}{\kilo\volt\per\meter}).}
\label{pdfFig.}
\end{center}
\end{figure}

A static electric field sensor is installed at the CODALEMA site, giving every 3 seconds the value of the static vertical component of the atmospheric electric field~$\xi$.
In normal conditions (fair weather), the value of the atmospheric electric field is around \SI{140}{\volt\per\metre}. In thunderstorm conditions, the absolute value can reach \SI{10^5}{\volt\per\metre} at ground level. The probability density function of the atmospheric electric field values is presented in Fig.~\ref{pdfFig.}.
For each of the 18 LF events listed in table~\ref{lfeventtable}, we know the atmospheric electric field value within less than 1.5~s of the time of the event; these values are displayed in red in Fig.~\ref{pdfFig.}, together with the number of LF events in the corresponding bin (bin size is \SI{1.4}{\kilo\volt\per\metre}). Seven of them correspond to thunderstorm conditions, i.e. outside of the main peak. Using the probability distribution function, we can compute the following probabilities, at the time of the LF event detection: $P(\xi>\xi_\text{atm}(t_\text{LFevent}))$ for positive values and $P(\xi<\xi_\text{atm}(t_\text{LFevent}))$ for negative values. These probabilities are shown in the 6$^\text{th}$ column of table~\ref{lfeventtable}. In order to check whether the atmospheric electric field values at the time of detection of the LF events are compatible or not with the global probability density function, we perform the Fischer combined probability test~\cite{10.2307/2681650,fischerbook}. We find that the $\chi^2_\text{data}$ associated to the 18 individual probabilities is 133.7. According to Fischer, this value is extracted from a $\chi^2$ law with $36$ degrees of freedom. The $p$-value of $\chi^2_\text{data}$ is $4.8\times 10^{-13}$.
The conclusion is that the high values of the atmospheric electric field at the time of detection of the LF events are not compatible with a random coincidence: the LF detection of cosmic rays is strongly favored by thunderstorm conditions.

As an example, let us consider one of the 18 LF events detected during storm conditions. During the day of the event, the atmospheric electric field had a chaotic behavior from 09:00 to 18:00, exhibiting large electric field values. Around the time of the event (11:28), the atmospheric electric field was equal to \SI{-12.3}{\kilo\volt\per\metre}, about $20~\sigma$ from the average value during normal conditions (\SI{140}{\volt\per\metre}). It is worth noticing that the event presented in Fig.~\ref{mse} was detected under normal atmospheric electric field conditions. Furthermore, any abnormal atmospheric electric field would strongly complicate the analysis comparing the observed MF signals to the simulated ones since the latter would require to perform shower simulations assuming a minimum knowledge of the atmospheric electric field profile as a function of altitude.

As already observed in the past~\cite{APEL20111295}, it is likely that the radio signal experiences an amplification due to the local atmospheric electric field, making it possible to be detected even for low energy showers. This amplification of the LF signal due to thunderstorm conditions could be at the origin of the large electric field values recorded at the time of the pioneer experiments though, apart in~\cite{castagnoli1991}, atmospheric electric field conditions are not mentioned, making it impossible to confirm this hypothesis.

\subsection{How to explain the non-observation of the SDP at Nan\c{c}ay?}

\begin{figure}[!ht]
\begin{center}
	\includegraphics[scale=0.3]{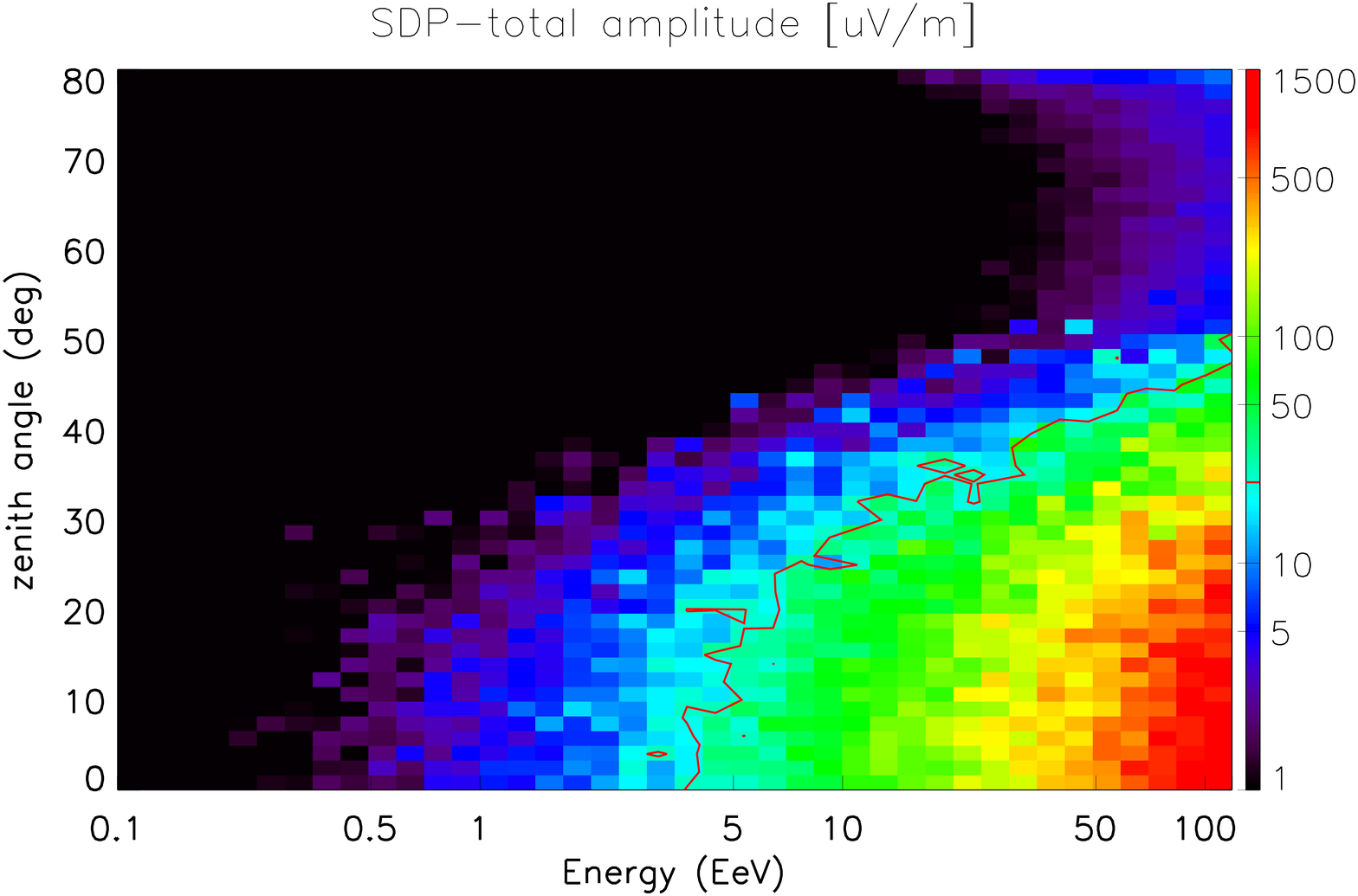}
	\includegraphics[scale=0.3]{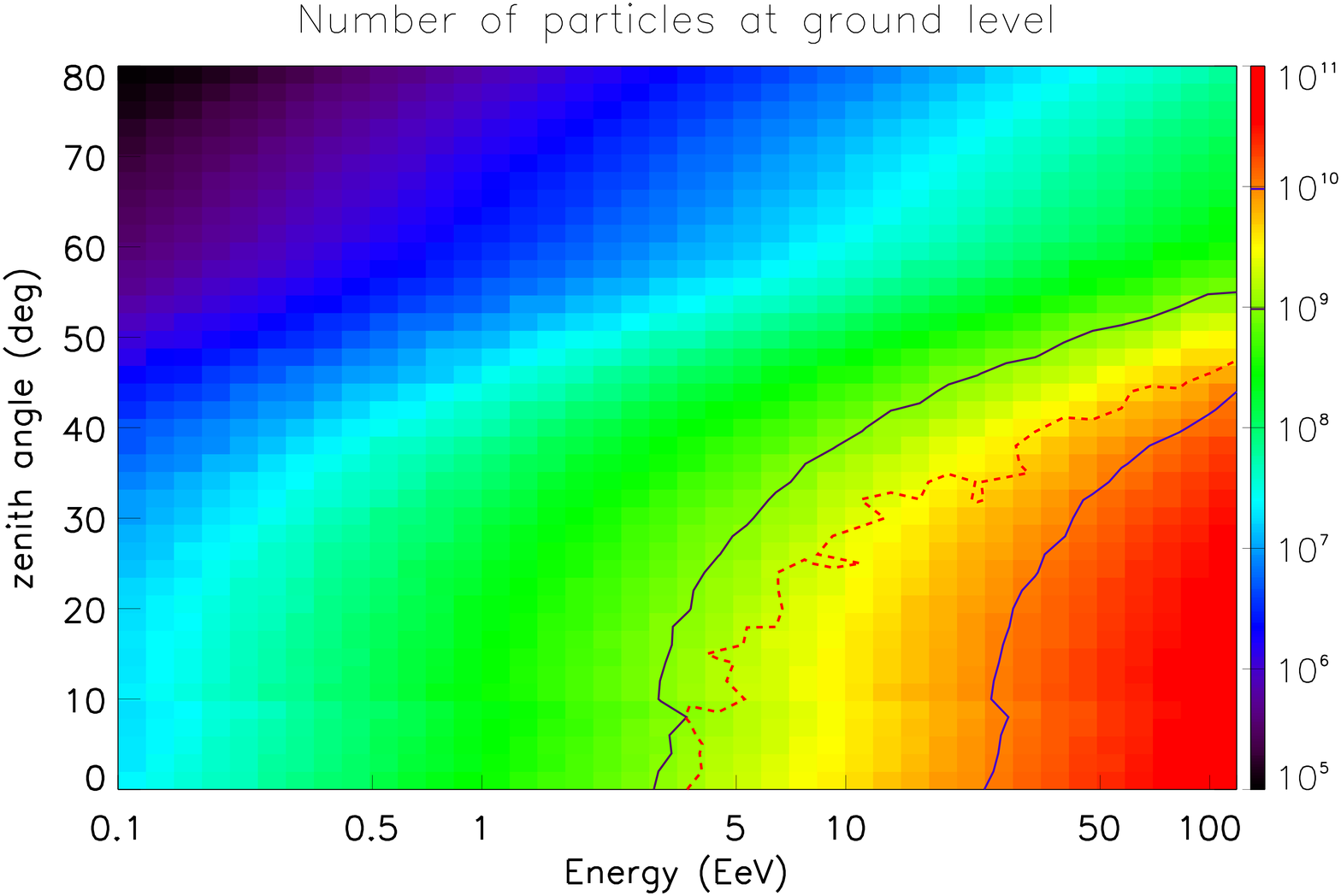}
	\caption{Top: SDP total amplitude as a function of the primary proton energy and shower zenith angle for the altitude of Nan\c{c}ay and for an antenna at \SI{200}{\metre} north of the shower core. Each bin contains the amplitude averaged over 5~showers. The frequency band is \SI{1-10}{\mega\hertz}. The contour corresponds to the detection threshold (\SI{23}{\micro\volt\per\metre}).
Bottom: number of particles reaching the ground at CODALEMA. Each bin contains the number averaged over~$7$~showers. The two solid contours correspond to~$10^9$ and~$10^{10}$~particles. The dashed contour corresponds to the detection threshold of~\SI{23}{\micro\volt\per\metre}.
}
\label{SDPground}
\end{center}
\end{figure}

\begin{figure}[!ht]
\begin{center}
	\includegraphics[width=14 cm]{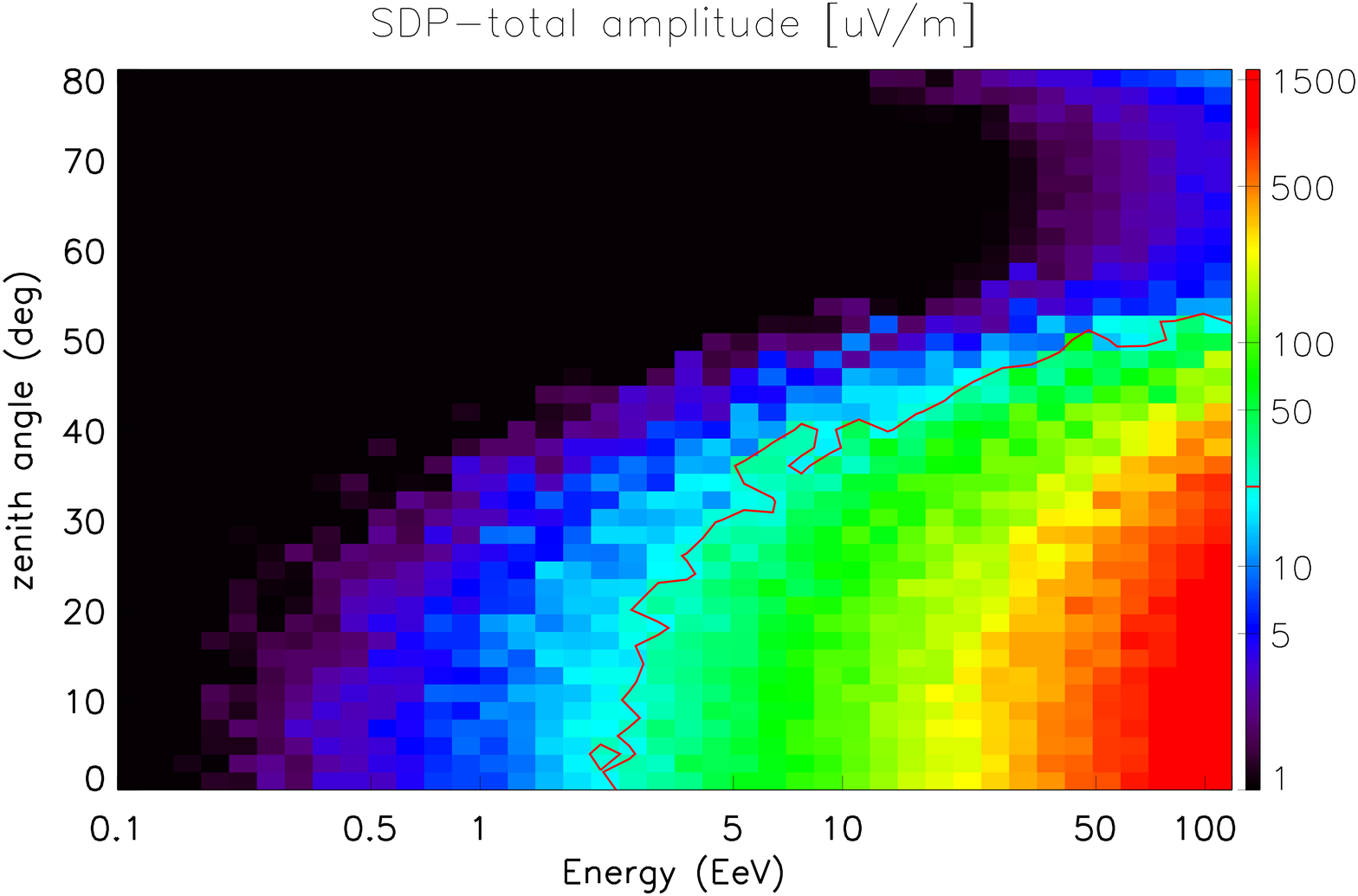}
	\includegraphics[width=12.5 cm]{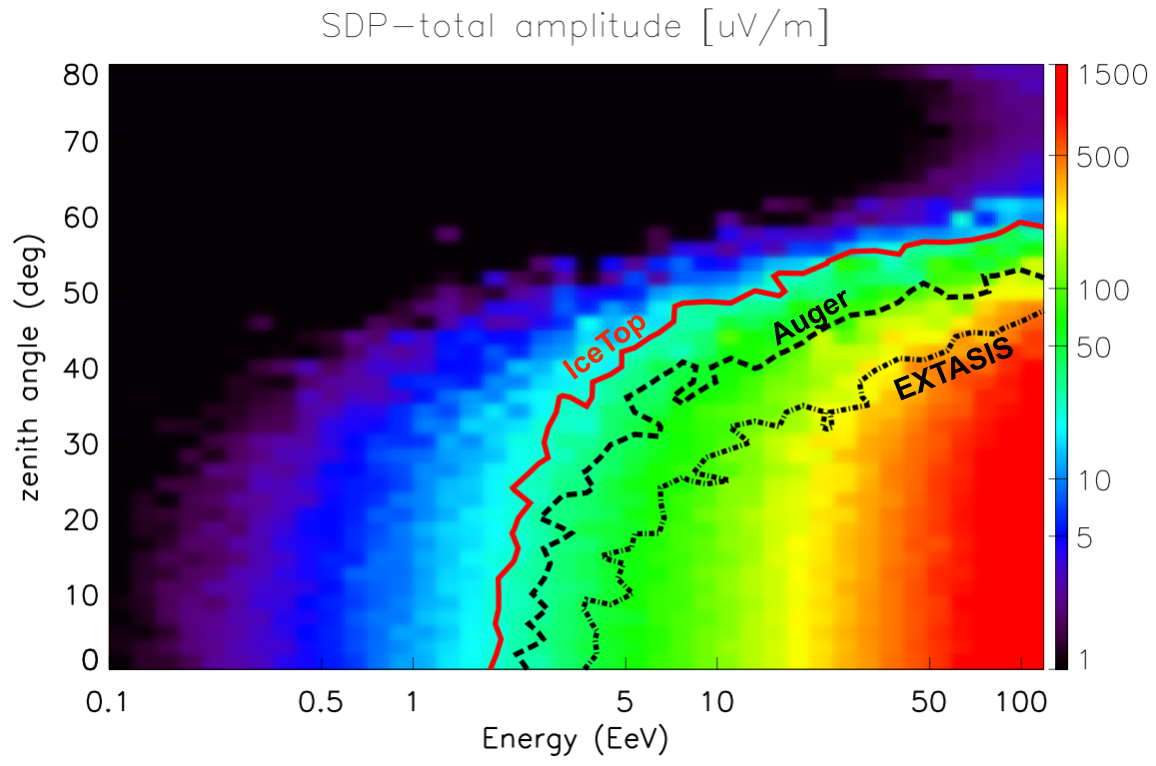}
	\caption{Top: same as Fig.~\ref{SDPground}-top but with a ground altitude of 1400~m corresponding to the altitude of the Pierre Auger Observatory site; the contour corresponds to the detection threshold of~{\SI{23}{\micro\volt\per\metre}}. Bottom: same figure for a ground altitude of 2800~m corresponding to the IceTop site. For an easier comparison, the contours obtained for the altitudes of EXTASIS (dot-dash line) and Auger (dash line) have been superimposed.}
\label{SDPgroundAuger}
\end{center}
\end{figure}

A strong contribution due to the sudden death of the shower could be obtained only if a lot of particles reach the ground, as shown in Fig.~\ref{SDPground}-bottom. This figure was obtained by simulating seven proton showers per bin of energy and zenith angle, assuming an altitude of \SI{130}{\metre} corresponding to the Nan\c cay site. At fixed energy, the number of particles reaching the ground decreases with increasing zenith angle. For example, for a primary energy around \SI{3\times 10^{18}}{\electronvolt}, and for vertical shower ($\theta\leqslant$\SI{10}{\degree}), the number of particles reaching the ground is of the order of $10^9$. We can infer that the sudden death signal for an event such as the one shown in the previous section ($E=\,$\SI{4\times10^{18}}{\electronvolt}, $\theta=\,$\SI{41}{\degree}) should not have been expected, because of a too small number of particles reaching the ground, estimated to be less than $6\times10^8$. This is confirmed by Fig.~\ref{SDPground}-top, featuring the total expected amplitude of the SDP for an antenna at \SI{200}{\metre} north of the shower core as a function of the primary energy and the shower zenith angle for the altitude of Nan\c{c}ay. Both figures are similar, corroborating the fact that the amplitude of the signal is directly related to the number of particles reaching the ground.
This analysis shows that the SDP could be detectable for a number of particles at ground level larger than $10^9$.
If we consider the detection threshold at low frequency as previously estimated, i.e. \SI{23}{\micro\volt\per\metre}, showers giving a detectable SDP at the altitude of Nan\c{c}ay should have an energy larger than \SI{4\times10^{18}}{\electronvolt} and a zenith angle smaller than the value indicated by the red dashed-line. We expect of the order of 0.3 shower per year having these characteristics at Nan\c{c}ay (assuming a duty cycle of 50~\% due to the day/night effect). This considerably hampers the possibility of detection and thus the confirmation of the existence of the sudden death phenomenon. However, as shown in Fig.~\ref{SDPgroundAuger}, the observation of the SDP could be significantly easier with LF antennas installed at places of higher altitudes such as the Pierre Auger Observatory (\SI{1400}{\metre}, Fig.~\ref{SDPgroundAuger}-top) or even better the IceTop site (2800~m, Fig.~\ref{SDPgroundAuger}-bottom). In Fig.~\ref{SDPgroundAuger}-bottom we display the \SI{23}{\micro\volt\per\metre} contour for IceTop but we also added the same contours for the altitudes of Auger (1400~m) and Nan\c{c}ay (130~m).
Going to higher altitudes implies a much larger number of particles at ground for showers of a few~EeV, which considerably increases the chances of observation of this phenomenon. 
We see that the IceTop site is better than Auger: at all zenith angles the energy threshold is smaller and with a differential flux decreasing as $\sim E^{-3}$ the number of detectable events is larger. For fixed area and observation time, we give in table~\ref{tabevents} the ratio of the number of detectable showers as a function of zenith angle.
\begin{table}[!ht]
\begin{center}
\begin{tabular}{|c|c|c|c|}\hline
zenith angle & $0^\circ$ & $30^\circ$ & $50^\circ$ \\ \hline
ratio IceTop/Auger  & 1.6 & 2.7 & 30 \\ \hline
ratio IceTop/EXTASIS & 4 & 16 & 342\\ \hline
ratio Auger/EXTASIS & 2.5 & 6 & 11.4\\ \hline
\end{tabular}
\caption{Ratio of the number of detectable events for the sites of IceTop, Auger and EXTASIS at fixed area and observation time as a function of zenith angle, assuming the same detection threshold of {\SI{23}{\micro\volt\per\metre}}.}\label{tabevents}
\end{center}
\end{table}

\begin{figure}[!ht]
\begin{center}
	\includegraphics[scale=0.2]{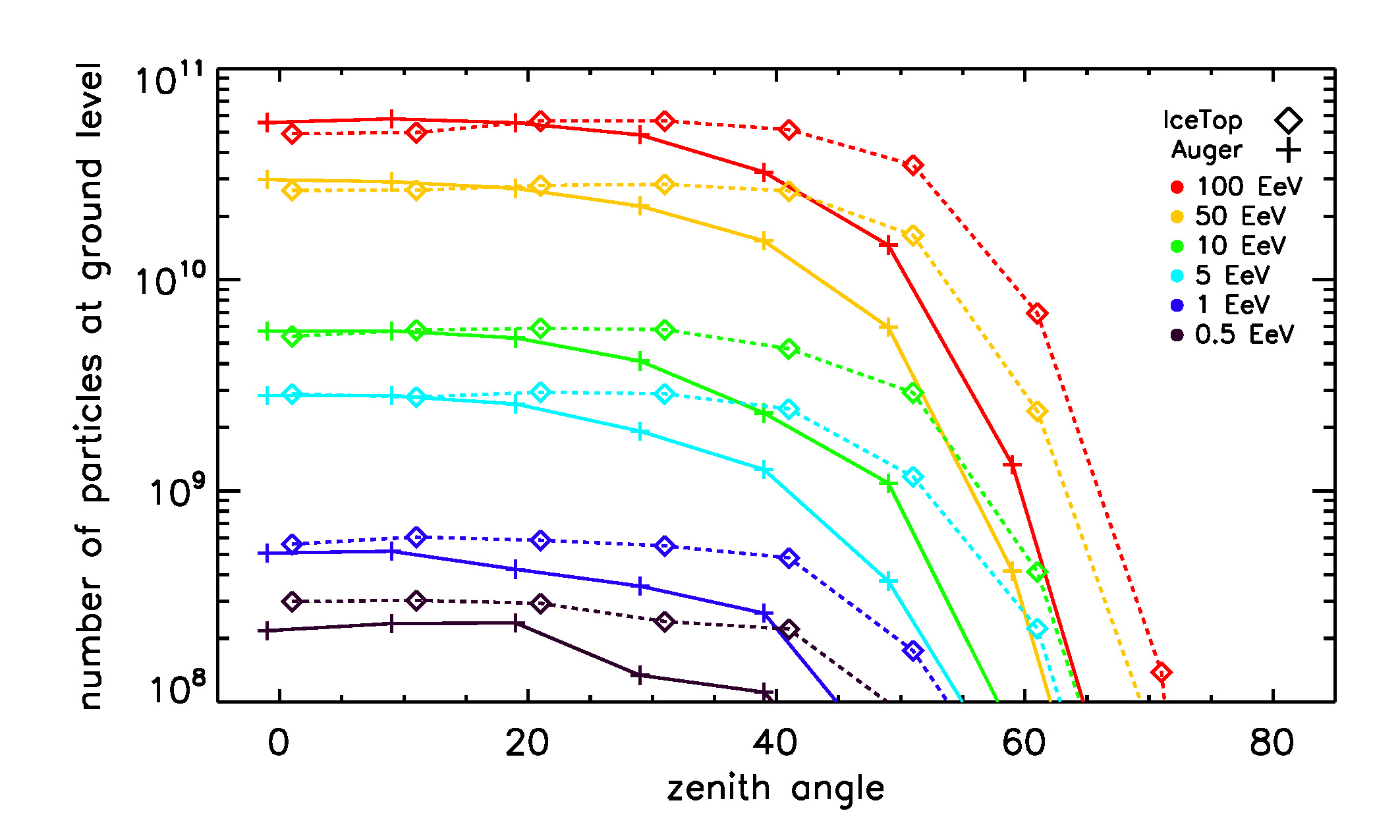}
	\caption{Number of particles at ground level as a function of the zenith angle, at various energies, for the Pierre Auger Observatory site (altitude 1400~m) and the IceTop site (altitude 2800~m).}
\label{ngroundtheta}
\end{center}
\end{figure}

These numbers are due to the evolution of the total number of secondary particles reaching the ground level as a function of the zenith angle and the observation site altitude. For instance, Fig.~\ref{ngroundtheta} presents this number as a function of the zenith angle at the Auger and IceTop sites for different energies.
If we take into account the zenithal acceptances of Fig.~\ref{SDPgroundAuger}-right, an effective detection area of 1~km$^2$, a duty cycle of 50~\% and an integration time of 1~year, then the number of showers that could be observed with the sudden death signal is: 0.33 at the altitude of EXTASIS, 0.89 at the altitude of Auger and 1.75 at the altitude of IceTop. We considered the spectral indices of the cosmic ray flux measured by the Telescope Array experiment given in~\cite{spindices}.
The size of IceTop and EXTASIS are roughly the same, 1~km$^2$, so that the search for the sudden death signal seems to be compromised. The Auger site is much more interesting as we can expect of the order of 2600~events per year (using 3000~km$^2$). A site at 3000~m of altitude and covering an area of 200~km$^2$ (such as GRAND~\cite{grand300}), would provide around 350~events per year.

\section{Conclusion and outlooks}

The EXTASIS experiment has detected several LF signals in correlation with cosmic-ray events. They have been seen in coincidence with CODALEMA, allowing for some of them to reconstruct the characteristics of the primary cosmic ray by combining MF signals with SELFAS3 simulations. Using these results, we have performed a precise simulation and compared it to the LF and MF data. While we expected an agreement for the comparison with the MF data, due to the fact that it is the standard band used for the radio reconstruction method, we have seen that the comparison with the LF data is also satisfactory. This is a very nice validation of the SELFAS3 code. This result shows, for the first time, the frequency spectrum of air showers measured over a large frequency range from 1 to \SI{200}{\mega\hertz}, despite the strength of the AM and FM bands in the Nan\c{c}ay vicinity. We have also confirmed that the detection range at low frequency is larger than in the classical band, as depicted in Fig.~\ref{MFLFrange}.

A search for LF events based only on the presence of low-frequency transients among the events recorded by EXTASIS, however, did not yield any evidence of events without a MF counterpart, confirming the conclusion that, when it exists, the low-frequency contribution of the radio signal of the atmospheric particle showers is much smaller and more difficult to detect than the contribution in the EMF band. Unfortunately, harsh atmospheric noise conditions hamper the detection at low frequency, for which the efficiency is already very poor. We have shown that the low rate of detection in the low-frequency band is mainly due to the atmospheric noise, which, in the best case, remains 10~times higher than the amplitude of the signal that we want to detect at the low altitude of Nan\c{c}ay; the duty cycle is around 50~\%, the noise being too high during night time. We also found a correlation with the atmospheric electric field, that probably amplifies the transient signal and lowers again the real detection efficiency in normal conditions. This amplification could explain at least part of the large shower electric field values recorded in the past, though this can not be definitely demonstrated a posteriori.

Concerning at last the observation of the expected SDP, our simulations show that the EXTASIS antennas, though sensitive enough regarding the LF noise conditions, are not installed on a favorable site.
Higher altitude sites are much more favorable: 2800~m (IceTop) would be better than 1400~m (Auger) but the size of the IceTop site provides only around 1 shower per year detectable with the SDP. Auger is well suited with its area of 3000~km$^2$ and could observe 2600~showers producing a detectable SDP per year. It would be very interesting to perform MHz measurements in the Auger-Horizon project~\cite{augerhorizon}, initially dedicated to the precise measurement of inclined showers in the usual range 30-80~MHz. Also the GRAND site, if confirmed at 3000~m of altitude, could be very well suited to the SDP search (around 350~events per year assuming an area of 200~km$^2$).

\section*{Acknowledgements}
We thank the R\'egion Pays de la Loire for its financial support of the Astroparticle group of Subatech and in particular for its contribution to the EXTASIS experiment, and the PNHE (Programme National Hautes Energies) from the French institutes IN2P3 and INSU for having also always supported the CODALEMA experiment, both financially and scientifically. We also thank O. Deligny for useful discussions on the statistical test of section~\ref{atmelec}. Lastly, we thank the technical support teams of the Nançay Radioastronomy Observatory for valuable assistance during and after the deployment of the EXTASIS antennas.


\renewcommand\bibname{References} 
\bibliographystyle{myunsrt} 
\bibliography{biblio} 


\end{document}